\documentclass[twocolumn,showpacs,preprintnumbers,amsmath,amssymb]{revtex4}

\newcommand{\vev}[1]{\left\langle #1 \right\rangle}
\begin{document}
\preprint{hep-ph/0506057}
\title{Five-dimensional Yang-Mills-Einstein supergravity on orbifolds:\\ Parity assignments}
\author{Sean McReynolds}
\email{sean@phys.psu.edu}
\affiliation{Physics Department\\ Pennsylvania State University\\ University Park, PA. 16802, USA}
\begin{abstract}
We discuss the options for parity assignments in (on-shell) $\mathcal{N}=2$ five-dimensional Yang-Mills-Einstein supergravity theories (YMESGTs) coupled to tensor and hypermultiplets on the orbifold spacetime $\mathcal{M}_{4}\times S^{1}/\mathbb{Z}_{2}$.  Along the lines of orbifold-GUTs, we allow for general breaking of the five-dimensional gauge group at the orbifold fixed points.  We then extend the discussion to the case where the orbifold is $\,S^{1}/(\mathbb{Z}_{2}\times\mathbb{Z}_{2})$.  We do not presume the existence of fields with support only at fixed points.  As in the familiar case of (rigid) super-Yang-Mills theories on such orbifolds, only bulk hypermultiplets can lead to chiral multiplets in complex representations of the gauge group on the boundaries.  Massless chiral multiplets coming from bulk vector or tensor multiplets can potentially be used as Higgs supermultiplets, though a ``doublet-triplet" splitting via parity assignments is not available for the tensor sector.  We also find parity assignments for objects other than fields that appear in the Lagrangian, which will partially determine the structure of interactions of the boundary theories.  Assigning odd parities to the scalar sector of vector/tensor multiplets requires the four-dimensional boundary moduli spaces to lie on the boundary of the classical K\"{a}hler cone, which corresponds to collapsed Calabi-Yau 2-cycles at the orbifold fixed points in a compactification of eleven-dimensional supergravity.  There is an ambiguity in how to effect odd parity for the field-independent $C_{IJK}$ tensor of the 5D theory, which may admit a classical interpretation as Calabi-Yau 4-cycles collapsing to either 2- or 0-cycles.\end{abstract}
\maketitle 
\section{Introduction}\label{sec:introduction}
Phenomenological field-theoretic model building has recently refocused on scenarios in which the universe appears higher-dimensional above some energy scale to examine new electroweak and supersymmetry breaking schemes~\cite{ewsusy} and new strong-electroweak unification scenarios~\cite{guts}.  In~\cite{Freund:1982pf}, it was pointed out that the estimated scale of strong-electroweak unification was around the energy scale where a Kaluza-Klein type universe may not be able to be approximated by a 4D theory, in which case grand unification would occur in higher dimensions.  In~\cite{Ant}, it was suggested that the size of an extra dimension could be much larger (TeV scale) within the framework of perturbative string theory (one of the motivations was to tie this scale to the $\mathcal{N}=1$ supersymmetry breaking scale).  Subsequently, the Horava-Witten (HW) scenario~\cite{HW} and Randall-Sundrum (RS) scenarios~\cite{RS} served as the most recent revival of interest in intermediate scale five-dimensional spacetimes in which the ground state spacetime is a singular space that is isomorphic to a spacetime with boundaries.  In particular, the ``orbifold" spacetime $\mathcal{M}_{4}\times S^{1}/\Gamma$ (upstairs picture) corresponds to a manifold with two boundaries separated by a 5D bulk (downstairs picture).  This spacetime can resolve a number of issues in the supersymmetric Standard Model and supersymmetric Grand Unified Theories (GUTs).

Since an orbifold spacetime is singular, field theories are not well-defined on it, which requires some further interpretation (in the downstairs picture, the boundary is sharp).  Supergravity admits solitonic solutions that could ultimately be interpreted as the boundaries of these theories; these solutions are domain walls with some thickness, smoothing out the singular nature of sharp boundaries.  Supergravity in orbifold scenarios is interesting for other reasons: as the familiar argument goes, the gauge couplings in supersymmetric versions of the Standard Model unify~\cite{sin2:91} at a scale close to the scale at which quantum gravitational effects are expected to be non-negligible.  Since there is a significant interpolation involved in these suggestive results, one might look for unification of gauge \textit{and} gravitational couplings.  Therefore, the next step in bottom-up model building is to consider supergravity versions of the interesting scenarios. 

Orbifold constructions have been performed many times in the literature for both rigidly and locally supersymmetric field theories; for examples of the former, see~\cite{ewsusy, guts}; for the latter, see~\cite{JB, offshell, YL:03dec, ZGAZ:04jul, AK04}.  The generic result is a theory with 4D boundaries with 1/2 supersymmetry and broken gauge group; the low energy description is a 4D effective theory for each boundary.  However, a systematic classification of the types of boundary conditions available via parity assignments has not been performed for Yang-Mills-Einstein supergravity theories (YMESGTs) coupled to vector, hyper-, and tensor multiplets (usually, orbifold supergravity theories are considered in the context of HW or (original) RS type scenarios in which the Standard Model gauge, and sometimes matter, fields are supported only on the boundaries).  In this paper, we aim to provide a more complete list of options, including a careful look at the tensor sector, for the low energy spectrum via parity assignments in the simple case of the $S^{1}/\mathbb{Z}_{2}$ orbifold.  We do not presume boundary-localized field content, which would follow from the appropriate coupling of 3-branes on the orbifold fixed-planes (or other massless fields arising in an M-theoretic framework).  We will then extend our results to the case $S^{1}/(\mathbb{Z}_{2}\times\mathbb{Z}_{2})$.  Some of the results are generic to theories with super-Yang-Mills coupled to hypermultiplets, while others are unique to supergravity.  While orbifold-GUTs are a main motivation, the results are not restricted to these scenarios.  As a novel example of 5D GUT in the framework of supergravity, we illustrate some of the parity assignments using an $SU(5,1)$ gauging.

In the next section, we will make generic remarks about field theory on orbifolds.  In section~\ref{sec:supermultiplets appearing in Gamma=Z2 case}, we list the $\mathcal{N}=1$ supermultiplets that have propagating modes on the orbifold fixed planes of the spacetime $M_{4}\times S^{1}/\mathbb{Z}_{2}$, as dictated by consistent $\mathbb{Z}_{2}$-parity assignments.  We first discuss the pure Yang-Mills-Einstein and hypermultiplet sectors in turn.  The tensor sector, discussed in section~\ref{sec:tensor multiplet couplings}, is subtler and has not been discussed in the literature much.  In section~\ref{sec:objects other than fields}, we list the $\mathbb{Z}_{2}$-parity assignments for objects other than fields appearing in the Lagrangian; in particular, spacetime- and field-independent quantities may be assigned odd parities, without the need to assume that there is a formulation of the theory in which an odd-parity field is responsible for this.  In section~\ref{sec:extension to Gamma=Z2xZ2}, we extend the results to the case in which the fifth dimension is $S^{1}/(\mathbb{Z}_{2}\times \mathbb{Z}_{2})$, which can be more phenomenologically interesting.  We end with conclusions (section~\ref{sec:conclusion}) and some future directions (section~\ref{sec:future directions}).  Appendix A contains some conventions used in this paper, while appendix B contains a discussion of the fermionic parity assignments.  

\section{Supergravity on $S^{1}/\Gamma$}\label{sec:supergravity on S1/Z2}

In modeling five-dimensional spacetimes with four-dimensional boundaries, we can choose a particular construction using a spacetime of the form $M_{4}\times S^{1}/\Gamma$, where $\Gamma$ is a discrete group that acts non-freely on the circle \footnote{We'll refer to this as an orbifold spacetime, even though orbifolds are usually defined as spaces with singularities of codimension greater than one.}.  
The orbifold $\mathcal{M}_{4}\times S^{1}/\Gamma$, which has two 4D fixed planes, is isomorphic to a groundstate spacetime $\mathcal{M}_{4}\times \mathcal{I}$, where $\mathcal{I}$ is a closed interval so that this is a spacetime with two disjoint 4D boundaries.  Instead of considering the 5D theories with these spacetime boundaries (the downstairs picture), it is often convenient to compactify the 5D theory on $S^{1}$, followed by assignment of $\Gamma$-parities to quantities in the theory (the upstairs picture).   Since the Lagrangian (density) must be even under $\mathbb{Z}_{2}$, there are constraints on the relative parities of the fields and other objects.  Further constraints are imposed by the consistency of local coordinate transformations, supersymmetry transformations and gauge transformations.  The choice of $\Gamma$ reflects different classes of boundary conditions from the downstairs point of view.  We will first consider the simplest case $\Gamma=\mathbb{Z}_{2}$, which results in a theory with equivalent spectra and interactions at the two $\Gamma$ fixed points.  

The choice of the way $\mathbb{Z}_{2}$ acts on quantities in the theory reflects a particular set of consistent boundary conditions.  The $\mathbb{Z}_{2}$ acts as reflections on the $S^{1}$ covering space $[-\pi R,\,\pi R]$ (where $\{-\pi R\}\equiv \{\pi R\}$), with fixed points at $\{0\},\{\pi R\}$.  To leave the space $\mathcal{M}_{4}\times S^{1}/\mathbb{Z}_{2}$ invariant under this action, the coordinate functions, basis vectors, basis 1-forms, and metric components have
\begin{gather}
P(x^{\mu}; \;\partial_{\mu}; \;dx^{\mu})=+1\;\;\;\;P(x^{5}; \;\partial_{5}; \;dx^{5})=-1\nonumber\\
P(\hat{g}_{\mu\nu}; \;\hat{g}_{\,55})=+1\;\;\;\;P(\hat{g}_{\,\mu5})=-1,\nonumber
\end{gather}
where $P(\Phi)$ denotes the $\mathbb{Z}_{2}$ parity of the object $\Phi$.  The fixed planes are non-oriented. 

Fields carrying internal quantum numbers are sections of a fiber bundle, with spacetime being the base space.  In such a situation, it makes sense for the action of $\mathbb{Z}_{2}$ to be lifted from the base space to the total space~\cite{orbifolds}.  There are a number of ways to perform this lift, corresponding to various classes of boundary conditions.  Just as the $\mathbb{Z}_{2}$ action on the covering space $S^{1}$ results in a singular space $S^{1}/\mathbb{Z}_{2}$, the $\mathbb{Z}_{2}$ action on the total space will, in general, change the structure of the fibers over the base space.  

Objects other than fields appearing in the Lagrangian generally carry representation indices of the gauge group.  Such quantities are structures appearing in the gauge bundle, and are therefore generally affected by modifications of the gauge bundle resulting after $\mathbb{Z}_{2}$ action (even if they are field independent).  This gives meaning to assigning these objects $\mathbb{Z}_{2}$ parities.                   

Although physical states on $\mathcal{M}^{4}\times S^{1}/ \mathbb{Z}_{2}$ must be even under $\mathbb{Z}_{2}$-action, the field operators can carry even or odd parity.  We will take the expansion of an odd parity field on the orbifold to have an $n$th term of the form  
\begin{equation}\begin{split}
\Phi^{(n)}_{-}(x^{\mu},x^{5})=&A_{n}\Phi^{(n)}_{-}(x^{\mu})\sin(n x^{5} /R)\\
&+B_{n}\Phi^{(n)}_{-}(x^{\mu})\epsilon(x^{5})\cos(n x^{5}/R),\label{phi}
\end{split}\end{equation}
where $\epsilon(x^{5})$ is $+1$ for $(-\pi R,0)$ and $-1$ for $(0,\pi R)$; the $\Phi^{(n)}_{-}(x^{\mu})$ are even; and $A_{n},\,B_{n}$ are normalization factors. 

To avoid $\delta^{\,2}$ factors in the Lagrangian ($\delta$ being the Dirac distribution), bosonic fields cannot have $\epsilon(x^{5})$ factors in the above expansion.  Therefore, \textit{we impose the condition $B_{n}=0$ for odd bosonic fields, which therefore vanish on the orbifold fixed planes}. 

On the other hand, the equations of motion and kinetic terms for fermionic fields involve a single derivative, so $\epsilon(x_{5})$ factors are allowed (they will give rise to $\delta(x_{5})$ factors in the equations of motion and upstairs picture Lagrangian).  Therefore, \textit{fermionic fields on $S^{1}/\mathbb{Z}_{2}$ are not well-defined on the upstairs picture fixed planes}.

Odd-parity field-independent objects $C^{I_{1}\ldots I_{n}}_{J_{1}\ldots J_{n}}$ carrying gauge indices can generally be redefined as either $\epsilon(x^{5})C^{I_{1}\ldots I_{n}}_{J_{1}\ldots J_{n}}$ or $\kappa(x^{5})C^{I_{1}\ldots I_{n}}_{J_{1}\ldots J_{n}}$, where $C^{I_{1}\ldots I_{n}}_{J_{1}\ldots J_{n}}$ now has even parity, and $\kappa(x^{5})$ is
\begin{equation} 
\kappa(x^{5})=\left\{ \begin{array}{cl}
0 & \mbox{for }\; x^{5}=-\pi R\\
-1 & \mbox{for }\; -\pi R< x^{5} < 0\\
0 & \mbox{for }\; x^{5}=0\\
+1 & \mbox{for }\; 0<x^{5}<\pi R
\end{array} \right. \label{kappa} \end{equation}      
However, consistency may require one or the other. 

\section{Supermultiplets appearing in $\Gamma=\mathbb{Z}_{2}$ case}\label{sec:supermultiplets appearing in Gamma=Z2 case}
\subsection{Pure Yang-Mills-Einstein supergravity}\label{sec:pure YMESGT}  

An $\mathcal{N}=2$ 5D Maxwell-Einstein supergravity theory (MESGT)~\cite{MESGT} consists of a minimal supergravity multiplet and $n_{V}$ vector supermultiplets.  The total field content is 
$$\{e^{\hat{m}}_{\hat{\mu}}, \Psi^{i}_{\hat{\mu}}, A^{I}_{\hat{\mu}}, \lambda^{i\,\tilde{p}}, \phi^{\tilde{x}}\},$$
where $I=(0,1,...,n_{V})$ labels the graviphoton and vector fields from the $n_{V}$ vector multiplets; $i=(1,2)$ is an $SU(2)_{R}$ index; and $\tilde{p}=(1,...,n_{V})$ and $\tilde{x}=(1,...,n_{V})$ label the fermions and scalars from the $n_{V}$ vector multiplets.  The scalar fields parametrize an $n_{V}$-dimensional real Riemannian manifold $\mathcal{M}_{R}$, so the indices $\tilde{p},\tilde{q},\ldots$ and $\tilde{x},\tilde{y},\ldots$ may also be
viewed as flat and curved indices of $\mathcal{M}_{R}$, respectively.  The supersymmetry parameters $\epsilon^{i}$, the gravitini $\Psi^{i}_{\mu}$, and the spin-1/2 fields $\lambda^{\tilde{p}\,i}$ are 5D symplectic-Majorana spinors (see appendix A), which can be written in 2-component spinor notation as
\[
\epsilon^{1}=\left( 
\begin{array}{c} 
\eta \\
e\zeta^{*}
\end{array}
\right)\;\;\;\; \epsilon^{2}=\left( 
\begin{array}{c} 
\zeta \\
-e\eta^{*}
\end{array}
\right)
\]  
\begin{equation}
\Psi^{1}_{\mu}=\left( 
\begin{array}{c} 
\alpha_{\mu} \\
e\beta^{*}_{\mu}
\end{array}
\right)\;\;\;\; \Psi^{2}_{\mu}=\left( 
\begin{array}{c} 
\beta_{\mu} \\
-e\alpha^{*}_{\mu}
\end{array}
\right)\label{5dfermions}
\end{equation}
\[
\lambda^{\tilde{p}\;1}=\left( 
\begin{array}{c} 
\delta^{\tilde{p}} \\
e\gamma^{\tilde{p}\;*}
\end{array}
\right)\;\;\;\; \lambda^{\tilde{p}\;2}=\left( 
\begin{array}{c} 
\gamma^{\tilde{p}} \\
-e\delta^{\tilde{p}\;*}
\end{array}
\right).
\]
Introducing $(n_{V}+1)$ parameters $\xi^{I}(\phi)$ depending on the scalar fields, we define a cubic polynomial
$\mathcal{V} (\xi)=C_{IJK}\xi^{I}\xi^{J}\xi^{K}$,
where $C_{IJK}$ is a constant rank-3 symmetric tensor.  This polynomial determines a symmetric rank-2 tensor 
$$a_{IJ}(\xi)=-\frac{1}{3}\frac{\partial}{\partial
\xi^{I}}\frac{\partial}{\partial \xi^{J}} \ln\mathcal{V} (\xi).$$  
The parameters $\xi^I$ can be interpreted as coordinate functions for an   
$(n_{V}+1)$-manifold, called the ambient space.  
The tensor $a_{IJ}$, which may have indefinite signature, defines a metric on
this space.  However, the coordinates are restricted via
$\mathcal{V} (\xi)>0$ so that the metric is positive definite, which means that
the manifold is Riemannian.  The equation $\mathcal{V} (\xi)=k$ ($k\in \mathbb{R}^{+}$) defines a
family of real
hypersurfaces, and in particular 
\begin{equation}\mathcal{V} (\xi)=1 
\label{V=1}\end{equation}
defines a real $n_{V}$-manifold corresponding to the scalar manifold $\mathcal{M}_{R}$.  The functions $h^{I}$ and $h^{I}_{\tilde{x}}$ that appear in fermionic terms of the Lagrangian and susy transformations are directly proportional to $\xi^{I}|_{\mathcal{V}=1}$ and $\xi^{I}_{,\tilde{x}}|_{\mathcal{V}=1}$, respectively; the $h^{I}$ are essentially embedding coordinates of $\mathcal{M}_{R}$ in the ambient space.  In~\cite{MESGT}, it was shown that, when (\ref{V=1}) holds, the $C_{IJK}$ may be put in the ``canonical form"
\begin{equation}\begin{split}
&C_{000}=1,\;\;C_{0ij}=-\frac{1}{2}\delta_{ij},\\
&C_{00i}=0,\;\;C_{ijk}=\mbox{arbitrary},\label{canonical}
\end{split}\end{equation}   
where $I=(0,i)$, $i=1,\ldots, n_{V}$.  We denote the restriction of the ambient space metric to $\mathcal{M}_{R}$ as
$$\stackrel{\circ}{a}_{IJ}=a_{IJ}|_{\mathcal{V} =1}.$$
The metric of the scalar manifold is then the pullback of the restricted ambient space metric to $\mathcal{M}_{R}$:
$$g_{\tilde{x}\tilde{y}}=\frac{3}{2}\stackrel{\circ}{a}_{IJ}h^{I}_{,\tilde{x}}h^{J}_{,\tilde{y}}.$$
Both of these metrics are positive definite due to the constraint
$\mathcal{V}>0$.

The $C_{IJK}$ tensor completely determines the MESGT Lagrangian~\cite{MESGT}.  Therefore, the global symmetry group of the Lagrangian is given by the symmetry group, $G$, of this tensor, along with automorphisms of the $\mathcal{N}=2$ superalgebra; that is, the symmetry group of the Lagrangian is $G\times SU(2)_{R}$.  Since $G$ consists of symmetries of the full Lagrangian, they are symmetries of the scalar sector in particular, and therefore isometries of the scalar manifold $\mathcal{M}_{R}$: $G\subset Iso(\mathcal{M}_{R})$ (the $SU(2)_{R}$ action is trivial on the scalars).  The full Lagrangian, however, is not necessarily invariant under the full group $Iso(\mathcal{M}_{R})$.

A subgroup $K\subset G$ may then be gauged if $n_{V}+1\geq \mbox{dim}[K]$.  The vector fields decompose into 
\[ \mathbf{n_{V}+1}=\mbox{\textbf{adj}}(K)\oplus \mbox{\textbf{non-singlets}}(K)\oplus \mbox{\textbf{singlets}}(K). \]
Under infinitesimal $K$-transformations parametrized by $\alpha^{I}$, the bosonic fields transform as:
\begin{equation}\begin{split}
\delta_{\alpha}A^{I}_{\hat{\mu}}&=-\frac{1}{g}\partial_{\hat{\mu}}\alpha^{I}+\alpha^{J}f^{I}_{JK}A^{K}_{\hat{\mu}}\\
\delta_{\alpha}\phi^{\tilde{x}}&=\alpha^{I}K^{\tilde{x}}_{I},\label{alpha}
\end{split}\end{equation} 
where $K^{\tilde{x}}_{I}$ are a set of $n_{V}$ Killing vectors on the scalar manifold parametrized by the $\phi^{\tilde{x}}$; and $f^{I}_{JK}$ and $\alpha^{I}$ vanish if any index corresponds to a spectator vector field~\footnote{If we gauge a compact group, there is always at least one singlet (spectator) vector field, which can be identified as the graviphoton.}.

Now the $C_{IJK}$ must be a rank-3 symmetric invariant of $K$.  If $K$ is compact, then only $C_{ijk}$ (see (\ref{canonical})) must be a rank-3 symmetric invariant. 
The 5D bosonic YMESGT Lagrangian is~\cite{MESGT}
\begin{equation}\begin{split}
&\hat{e}^{-1}\mathcal{L}_{bos}=\\
&-\frac{1}{2}\hat{R}
-\frac{1}{4}\stackrel{\circ}{a}_{IJ}\mathcal{F}^{I}_{\hat{\mu}\hat{\nu}}\mathcal{F}^{J\,\hat{\mu}\hat{\nu}}
-\frac{1}{2}g_{\tilde{x}\tilde{y}}\,D_{\hat{\mu}}\phi^{\tilde{x}}D^{\hat{\mu}}\phi^{\tilde{y}}\\
&+\frac{\hat{e}^{-1}}{6\sqrt{6}}C_{IJK}\epsilon^{\hat{\mu}\hat{\nu}\hat{\rho}\hat{\sigma}\hat{\lambda}}
\{ F^{I}_{\hat{\mu}\hat{\nu}}F^{J}_{\hat{\rho}\hat{\sigma}}A^{K}_{\hat{\lambda}} \\
&\;\;\;\;\;\;\;\;\;\;\;\;\;\;\;\;\;\;\;\;\;\;\;\;\;\;\;\;\;\;\,+\frac{3}{2}gF^{I}_{\hat{\mu}\hat{\nu}}A^{J}_{\hat{\rho}}(f^{K}_{LM}A^{L}_{\hat{\sigma}}A^{M}_{\hat{\lambda}})\\
&\;\;\;\;\;\;\;\;\;\;\;\;\;\;\;\;+\frac{3}{5}g^{2}(f^{J}_{GH}A^{G}_{\hat{\nu}}A^{H}_{\hat{\rho}})(f^{K}_{LF}A^{L}_{\hat{\sigma}}A^{F}_{\hat{\lambda}})A^{I}_{\hat{\mu}} \} , 
\end{split}\end{equation} 
where hats indicate five-dimensional quantities and $\hat{e}$ is the determinant of
the f\"{u}nfbein.  The $A^{I}_{\mu}$ are abelian vector
fields and~\footnote{$A_{[\alpha}B_{\beta]}\equiv\frac{1}{2}(A_{\alpha}B_{\beta}-A_{\beta}B_{\alpha})$} 
\[\begin{split} F^{I}_{\hat{\mu}\hat{\nu}}&\equiv 2\partial_{[\hat{\mu}}A^{I}_{\hat{\nu}]}\\
\mathcal{F}^{I}_{\hat{\mu}\hat{\nu}}&\equiv F^{I}_{\hat{\mu}\hat{\nu}} + 2g A^{I}_{[\hat{\mu}}A^{J}_{\hat{\nu}]}\\
D_{\hat{\mu}}\phi^{\tilde{x}}&\equiv \partial_{\hat{\mu}}\phi^{\tilde{x}}+gK^{\tilde{x}}_{I}A^{I}_{\hat{\mu}}. \end{split}\]
The supersymmetry transformations of the bosonic fields are 
\begin{equation} \begin{split}
\delta\hat{e}^{\hat{m}}_{\hat{\mu}}=&\frac{1}{2}\bar{\epsilon}^{i}\Gamma^{m}\Psi_{\mu\;i}\\
\delta A^{I}_{\hat{\mu}}=&-\frac{1}{2}h^{I}_{\tilde{p}}\bar{\epsilon}^{i}\Gamma_{\mu}\lambda^{\tilde{p}}_{i}+\frac{i\sqrt{6}}{4}h^{I}\hat{\bar{\Psi}}^{i}_{\hat{\mu}}\epsilon_{i}\\
\delta \phi^{\tilde{x}}=&\frac{i}{2}f^{\tilde{x}}_{\tilde{p}}\bar{\epsilon}^{i}\lambda^{\tilde{p}}_{i},
\end{split} \label{susytrs}\end{equation}
where the $f^{\tilde{x}}_{\tilde{p}}$ are $n_{V}$-bein of the real scalar manifold $\mathcal{M}_{V}$.\vspace{1mm}\\ 
\textbf{Dimensional reduction of 5D $\mathcal{N}=2$ YMESGT}

In the ``upstairs" orbifold construction, one starts with a 5D theory, and compactifies on $S^{1}$.  It is sufficient for our purposes to use the dimensionally reduced theory, consisting of those fields satisfying $\partial_{5}\Phi=0$.  The dimensional reduction breaks the 5D local Lorentz invariance to a 4D local Lorentz
invariance.  The four local symmetries that are broken can be used to fix four
degrees of freedom in the f\"{u}nfbein.  Splitting $\hat{\mu}=(\mu, 5)$, we choose the parametrization for the f\"{u}nfbein to be~\cite{MESGT}  
\[ \hat{e}^{\hat{m}}_{\hat{\mu}}= \left( \begin{array}{ccc}
e^{-\frac{\sigma}{2} }e^{m}_{\mu}  & & 2 e^{\sigma}  C_{\mu} \\
0 & & e^{\sigma}
\end{array} \right). \]\vspace{4mm}
Since $\hat{g}_{\hat{\mu}\hat{\nu}}=\hat{e}^{\hat{m}}_{\hat{\mu}}\,\hat{e}^{\hat{n}}_{\hat{\nu}}\,\eta _{\hat{m}\hat{n}}$, 
\begin{equation} \begin{split}
\hat{g}_{\mu\nu}=&e^{-\sigma}g_{\mu\nu}+4e^{2\sigma}C_{\mu}C_{\nu}\\
\hat{g}_{5\,5}=&e^{2\sigma}\\
\hat{g}_{\mu\,5}=&2e^{2\sigma}C_{\mu}.
\end{split} \label{metric} \end{equation}
Furthermore, let $A^{I}_{\hat{\mu}}=(A^{I}_{\mu},\,A^{I}).$
Under infinitesimal local coordinate transformations of the compact
coordinate parameterized by $\xi^{5}(x^{\mu})$, the 4D fields
$A^{I}_{\mu}$ and $C_{\mu}$ transform as 
\begin{equation}\begin{split}
\delta_{\xi^5} A^{I}_{\mu}&=-\partial_{\mu}\xi^{5}A^{I} \\
\delta_{\xi^5} C_{\mu}&=-2\partial_{\mu}\xi^{5},\label{KKtrs}
\end{split}\end{equation}
with the remaining 4D bosonic fields being invariant.  One can interpret $\xi^{5}(x^{\mu})$ as a parameter for local $U(1)$ transformations, for which $C_{\mu}$ is a gauge field.  Note that the vector fields $A^{I}_{\mu}$ transform non-trivially under these $U(1)$ transformations.  To properly dimensionally reduce the theory, one must make field redefinitions in order to obtain $U(1)$ (KK)-invariant fields 
\[A^{I}_{\mu}\rightarrow A^{I}_{\mu}+2C_{\mu}A^{I}.\]
The dimensionally reduced bosonic Lagrangian becomes~\cite{GMZ05a}\begin{widetext}
\begin{equation} \begin{split}
e^{-1}\mathcal{L}_{DR} = 
&-\frac{1}{2}R
-\frac{3}{4}\stackrel{\circ}{a}_{IJ}D_{\mu}\tilde{h}^{I} D^{\mu}\tilde{h}^{J}
-\frac{1}{2}e^{-2\sigma}\stackrel{\circ}{a}_{IJ}D_{\mu}A^{I}D^{\mu}A^{J}\\
&-(\frac{1}{2}e^{3\sigma}+ e^{\sigma}\stackrel{\circ}{a}_{IJ}A^{I}A^{J})C_{\mu\nu}C^{\mu\nu}
-\frac{1}{4} e^{\sigma}\stackrel{\circ}{a}_{IJ}\mathcal{F}^{I}_{\mu\nu}\mathcal{F}^{\mu\nu\,J}
-e^{\sigma}\stackrel{\circ}{a}_{IJ}A^{I}\mathcal{F}^{J}_{\mu\nu}C^{\mu\nu}\\
&+\frac{e^{-1}}{2\sqrt{6}}C_{IJK}\epsilon^{\mu\nu\rho\sigma}(A^{I}\mathcal{F}^{J}_{\mu\nu}\mathcal{F}^{K}_{\rho\sigma}
+2A^{I}A^{J}\mathcal{F}^{K}_{\mu\nu} C_{\rho\sigma}
+\frac{4}{3}A^{I}A^{J}A^{K}C_{\mu\nu}C_{\rho\sigma})\\
&-g^{2}P,
\end{split} \end{equation}\end{widetext}
where $\tilde{h}^{I}\equiv e^{\sigma} h^{I}$, and
\begin{equation}
P\equiv \frac{3}{4}e^{-3\sigma}\stackrel{\circ}{a}_{IJ}(A^{K}f^{I}_{KL}h^{L})(A^{M}f^{J}_{MN}h^{N}),\label{YMESGTpot}
\end{equation}
and 
\begin{eqnarray}
D_{\mu}A^{I} & \equiv  & \partial_{\mu} A^{I} +g A_{\mu}^{J}f_{JK}^{I}A^{K}\\
D_{\mu}\tilde{h}^{I} & \equiv & \partial_{\mu} \tilde{h}^{I} + g A_{\mu}^{J}f_{JK}^{I}\tilde{h}^{K}\\
C_{\mu\nu} & \equiv & 2\partial_{[\mu}C_{\nu]}. 
\end{eqnarray}
Just as $A^{I}_{\mu}$ was redefined to be KK-invariant, we make the further redefinitions 
\begin{eqnarray}
\Psi^{i}_{\mu} & \rightarrow & \Psi^{i}_{\mu}+\Psi^{i}_{\dot{5}}C_{\mu}\\
\Gamma_{\mu} & \rightarrow & \Gamma_{\mu}+\Gamma_{\dot{5}}C_{\mu},
\end{eqnarray}
so that $\Psi^{i}_{\mu}$ and $\Gamma_{\mu}$ are now KK-invariant.  
The dimensionally reduced susy transformations of the bosonic fields are
\begin{equation}\begin{split}
\delta' e^{m}_{\mu}=\;&\frac{1}{2}\bar{\epsilon}^{i}\Gamma^{m} \Psi^{(4)}_{\mu\,i}\\
\delta A^{I}_{\mu}=\;&-\frac{1}{2}h^{I}_{\tilde{p}}\bar{\epsilon}^{i}\Gamma_{\mu}\lambda^{\tilde{p}}_{i}+\frac{1}{2}ie^{-\sigma}\bar{\Psi}^{i}_{\mu}\epsilon_{i}(\frac{\sqrt{6}}{2}\tilde{h}^{I}+A^{I})\\
\delta A^{I}=\;&-\frac{1}{2}h^{I}_{\tilde{p}}\bar{\epsilon}^{i}\Gamma_{\dot{5}}\lambda^{\tilde{p}}_{i}+\frac{\sqrt{6}}{4}i\bar{\psi}^{i}\epsilon_{i}h^{I}\\
\delta e^{\sigma}=\;&\frac{1}{2}\bar{\epsilon}^{i}\Gamma^{5}{\psi}_{i}\\
\delta C_{\mu}=\;&\frac{1}{2}e^{-\sigma}\bar{\epsilon}^{i}\Gamma^{5}\Psi_{\mu\,i}\\
\delta \phi^{x}=\;&\frac{i}{2}f^{x}_{a}\bar{\epsilon}^{i}\lambda^{a}_{i},
\end{split} \end{equation} 
where $\Psi^{i}_{\hat{\mu}}=(\Psi^{i}_{\mu},\psi^{i})$; $\delta'$ denotes the ``bare" susy transformation from five dimensions plus a local Lorentz transformation to maintain the condition $\hat{e}^{m}_{\dot{5}}=0$; and we have identified the four-dimensional gravitini to be 
\begin{equation}
\Psi^{(4)}_{\mu\,i}\equiv e^{n}_{\mu}\{\Psi_{n\,i}+\frac{1}{2}(\Gamma^{n})^{-1}\Gamma^{5}\Psi_{5\,i}\}.
\end{equation}
The dimensionally reduced Lagrangian can be written in terms of a K\"{a}hler scalar manifold with complex scalars   
\begin{equation}
z^{I}\equiv \frac{1}{3}\,A^{I}+\frac{i}{\sqrt{2}}\,\tilde{h}^{I}.\label{z}
\end{equation} 
More details of the dimensional reduction to 4D $\mathcal{N}=2$ supergravity can be found in~\cite{GMZ05a}.\vspace{1mm}\\
\textbf{Boundary propagating multiplets}

It's clear from appendix B that the action of $\mathbb{Z}_{2}$ on the supersymmetry spinors $\epsilon^{i}$ necessarily requires half of the components to be odd, so that the original eight supersymmetry currents will be broken to four on the boundaries, so that there is at most $\mathcal{N}=1$ susy there.  In terms of symplectic-Majorana spinors $\epsilon^{i}$, the $\mathbb{Z}_{2}$ action is represented as 
\[-i\Gamma^{5}\epsilon^{1}\;\;\;\;\mbox{and}\;\;\;\; i\Gamma^{5}\epsilon^{2}.\]
(The 4-component eigenspinors of the $\mathbb{Z}_{2}$ action are linear combinations of the two symplectic-Majorana spinors.)
The zero mode susy parameters can be written in the \textit{upstairs} picture as (see table~\ref{tab:Table17} in appendix B)
\[ \epsilon^{1}=\left( 
\begin{array}{c} 
\eta \\
\epsilon(x^{5})e\zeta^{*}
\end{array}
\right),\;\;\;\;\;\; \epsilon^{2}=\left( 
\begin{array}{c} 
\epsilon(x^{5})\zeta \\
-e\eta^{*}
\end{array}
\right), \]  
and so don't have a well-defined limit on the fixed-planes.  In the downstairs picture, on the other hand, fermions will have a well-defined limit at the boundaries (see~\cite{JB} e.g.); the fields in (\ref{5dfermions}) can be written at the boundaries either as left-chiral fermions with their right-chiral conjugates:
\[ \lambda^{\tilde{p}\;1}=\left( 
\begin{array}{c} 
\delta^{\tilde{p}} \\
0
\end{array}
\right),\;\;\;\;\;\; \lambda^{\tilde{p}\;2}=\left( 
\begin{array}{c} 
0 \\
-e\delta^{\tilde{p}\;*}
\end{array}
\right), \]
or right-chiral fermions with their left-chiral conjugates, which we denote with a bar:
\[ \bar{\lambda}^{\tilde{p}\;1}=\left( 
\begin{array}{c} 
0 \\
e\gamma^{\tilde{p}\;*}
\end{array}
\right),\;\;\;\;\;\; \bar{\lambda}^{\tilde{p}\;2}=\left( 
\begin{array}{c} 
\gamma^{\tilde{p}} \\
0
\end{array}
\right). \]
 
Assuming for simplicity that the only boundary-propagating vector fields are gauge fields for the bulk gauge group $K$, consistent parity assignments in the upstairs picture allow the following boundary multiplets in the downstairs picture:
\begin{center}\begin{tabular}{|ccc|}\hline
Multiplet & Representation & Type\\\hline 
$\{g_{\mu\nu},\;\Psi_{\mu}\}$ & $\mbox{$K_{\alpha}$ singlet}$ & \\ 
$\{A^{\alpha}_{\mu},\;\lambda^{\rho\,i}\}$ & $\mbox{adj}[K_{\alpha}]$ & $\mathbb{R}$ \\ 
$\{\bar{\lambda}^{p\,i},\;z^{a}\}$ & $R_{V}[K/K_{\alpha}]$ & $\mathbb{R}$ \\ 
$\{\Psi_{\dot{5}},\;z^{0}\}$ & $\mbox{$K_{\alpha}$ singlet}$ & \\ \hline
\end{tabular}\end{center}
where we've split the index $I=(0,\alpha, a)$; $\tilde{x}=(x,\chi)$; and $\tilde{p}=(p,\rho)$.  We have denoted the surviving gauge group on the boundaries as $K_{\alpha}$, and $R_{V}[K/K_{\alpha}]$ is the representation of the coset space elements.  The value of $n'$ in $\alpha=1,\ldots,n'$ and $a=(n'+1),\ldots, (n_{V}+1)$ is arbitrary for now, so there is complete freedom in choosing parities for vector multiplets; we will consider this more carefully in section~\ref{sec:constraints on assigning parities}.  The second to last multiplet consists of a chiral multiplet in a real representation and its CPT conjugate.  The case in which there are $K$-singlet vector fields propagating on the boundaries is straightforward.  

What happens when a non-compact group is gauged in five dimensions?  If the non-compact gauge fields were assigned \textit{even} parity, then a non-compact gauge group would appear in the 4D theory.  However, there would not be the proper degrees of freedom to give a ground state with compact gauge symmetry since the scalar degrees of freedom $A^{I}$ needed to form massive $\mathcal{N}=1$ vector multiplets must have odd parity.  Therefore, \textit{the non-compact gauge fields must be assigned odd parity}.  We will then get $\mathcal{N}=1$ chiral multiplets in the coset $K/H$, with $H$ the maximal compact subgroup of $K$.  This is a novel way of obtaining a 4D Higgs sector, along the lines of previous Higgs-gauge unifications in higher dimensions~\cite{gaugehiggs}, since there will be a distinct scalar potential for these scalars. \vspace{2mm}\\   
\subsection{Hypermultiplet couplings~\cite{general coupling}}\label{sec:hypermultiplet couplings}

A colection of $n_{H}$ hypermultiplets in five dimensions consist of $2n_{H}$ fermions and $4n_{H}$ real scalars, the latter parametrizing a quaternionic $n_{H}$-manifold $\mathcal{M}_{Q}$ with tangent space group $USp(2n_{H})\times SU(2)_{R}$.  We write the hypermultiplets as 
\[\{\zeta^{A},q^{\tilde{X}}\},\]
where $\tilde{X}=1,\ldots,4n_{H}$ are the curved indices of $\mathcal{M}_{Q}$; and $A=1,\ldots, 2n_{H}$ are flat, $USp(2n_{H})$ indices.  The $4n_{H}$-bein $f^{\tilde{X}}_{iA}$ relate scalar manifold curved and flat space metrics 
\[g_{\tilde{X}\tilde{Y}}f^{\tilde{X}}_{iA}f^{\tilde{Y}}_{jB}=\epsilon_{ij}C_{AB},\]
where $i,j=1,2$ are $SU(2)_{R}$ indices.  Note that, in contrast to the case of vector multiplets, the scalars form $2n_{H}$ $SU(2)_{R}$-doublets, while the $2n_{H}$ fermions are $SU(2)_{R}$-singlets.\footnote{In this paper, we are not considering gaugings of $SU(2)_{R}$ or its subgroups.}  In 2-component spinor notation, we write the fermions as 
\begin{equation}\zeta^{A}=\left( 
\begin{array}{c} 
\zeta^{A}_{1} \\
\zeta^{A}_{2}
\end{array}
\right). \label{5dfermions2}\end{equation}
If non-trivial isometries of $\mathcal{M}_{Q}$ are gauged, they act on the scalars as
\begin{equation}
\delta_{\alpha} q^{\tilde{X}}=\alpha^{I}K^{\tilde{X}}_{I}(q),\label{gaugeq}
\end{equation}
where the $K^{\tilde{X}}_{I}$ are the Killing fields on the quaternionic scalar manifold; and $\alpha^{I}$ are the same local transformation parameters as in the pure YMESGT sector.
The susy transformations for the scalars are 
\begin{equation} 
\delta q^{\tilde{X}}=-i\bar{\epsilon}^{i}\zeta^{A}f^{\tilde{X}}_{iA}.\label{susyq}
\end{equation}
The 5D bosonic hypermultiplet Lagrangian (coupled to a YMESGT) is 
\begin{equation}
\hat{e}^{-1}\mathcal{L}_{hyper}=-\frac{1}{2}g_{\tilde{X}\tilde{Y}}\mathcal{D}_{\hat{\mu}}q^{\tilde{X}}\mathcal{D}^{\mu}q^{\tilde{Y}}-2g^{2}V_{i\,A}V^{i\,A},\label{potq}
\end{equation}
where 
\[V^{i\,A}\equiv \frac{\sqrt{6}}{4}h^{I}K^{\tilde{X}}_{I}f^{A\,i}_{\tilde{X}}\]
and the $K$-covariant derivative is
\[D_{\mu}q^{\tilde{X}}\equiv \mathcal{D}_{\mu}q^{\tilde{X}}+gA^{I}_{\hat{\mu}}K^{\tilde{X}}_{I}(q),\]
where $\mathcal{D}_{\mu}$ is the covariant derivative associated with the tangent space Lorentz and $USp(2n_{H})\times SU(2)$ connections.  The dimensional reduction of the bosonic Lagrangian is straightforward and will not be quoted.\vspace{1mm}\\
\textbf{Boundary propagating multiplets}

Half of the hypermultiplet field content is required to have odd parity; therefore, let's split the index $\tilde{X}=(X,\chi)$, with $X=1,\ldots, 2n_{H}$ and $\chi=2n_{H}+1,\ldots, 4n_{H}$.  We let $q^{X}$ be the even parity fields, and $q^{\chi}$ the odd fields.  Similarly, we spit the index $A=(n,\tilde{n})$ with $n=1,\ldots, n_{H}$ and $\tilde{n}=n_{H}+1,\ldots, 2n_{H}$.
If we couple hypermultiplets in the quaternionic-dimensional representation $R_{H}[K]$ of the gauge group to a 5D YMESGT, the multiplets with boundary propagating modes will be
\begin{center}\begin{tabular}{|ccc|}\hline
Multiplet & Representation & Type\\\hline
$\{\zeta^{n}_{1},\,q^{X_{1}}\}$ & $R_{H}[K_{\alpha}]$ & $\mathbb{R}$ or $\mathbb{C}$\\
$\{\zeta^{\tilde{n}}_{2},\,q^{X_{2}}\}$ & $\overline{R}_{H}[K_{\alpha}]$ & $\mathbb{R}$ or $\mathbb{C}$\\\hline  
\end{tabular}\end{center}
where we have further split $X=(X_{1},X_{2})$ with $X_{1}=1,\ldots, n_{H}$ and $X_{2}=n_{H}+1,\ldots, 2n_{H}$. 
We get a left-chiral multiplet and its CPT conjugate.  Here $R_{H}[K_{\alpha}]$ is the decomposition of $R_{H}[K]$ under the group $K_{\alpha}\subset K$, and is the real-dimensional representation (see appendix A for conventions).\vspace{1mm}\\
\textbf{Example}

Consider the ``unified" MESGT with $SU(5,1)$ global symmetry group~\cite{GZ03, M05a}, whose vector fields are in 1-1 correspondence with the traceless elements of the Lorentzian Jordan algebra $J^{\mathbb{C}}_{(1,5)}$~\cite{GZ03}.  The theory is ``unified" in the sense that all of the vector fields of the 5D theory (including the bare graviphoton) furnish the $\mbox{adj}[SU(5,1)]$.  The $C_{IJK}$ tensor is a rank-3 symmetric invariant of the global symmetry group, so its components are proportional to the $d$-symbols of $SU(5,1)$.  

As in~\cite{M05a}, we can now couple hypermultiplets whose scalars parametrize the quaternionic manifold
\begin{equation}
\mathcal{M}_{Q}=\frac{E_{7}}{SO(12)\times SU(2)}\label{mfd}
\end{equation}
to the MESGT based on $J^{\mathbb{C}}_{(1,5)}$, gauging the common $SU(5,1)$ subgroup.  Then the five-dimensional ground state would have at most an $SU(5)\times U(1)$ gauge group coupled to hypermultiplets in the $\mathbf{1}\oplus\mathbf{5}\oplus\mathbf{10}$.  We may then make parity assignments by associating each type of index with an $SU(5)$ representation:
\begin{center}\begin{tabular}{|c|c|c|c|c|}\hline
$\alpha$ & $a$ & $0$ & $n$ & $\tilde{n}$\\\hline
$\,\mbox{adj}[SM]\,$ & $\,SU(5)/SM \oplus \mathbf{5}\oplus\mathbf{\bar{5}}\,$ & $\,\mathbf{1}\,$ & $\,\mathbf{1}\oplus\mathbf{\bar{5}}\oplus\mathbf{10}\,$ & $c.c.$ \\\hline
\end{tabular} \end{center}
The 4D low energy effective theory of each boundary will have an $\mathcal{N}=1$ supergravity multiplet; SM gauge multiplets; weak doublet and color triplet chiral multiplets both with a scalar potential; and left-chiral matter multiplets (including a sterile fermion multiplet) along with their right-chiral conjugates.  There are also the generic singlet left- and right-chiral multiplets coming from the 5D supergravity multiplet, as well as chiral multiplets in the symmetric space $SU(5)/SM$.\\
\textbf{Remark:}\\
At least within the category of homogeneous quaternionic manifolds, all occurences of hypermultiplets in the $\mathbf{5}$ and $\mathbf{10}$ of $SU(5)$ come from spaces admitting an $SO(10)$ isotropy subgroup under which these $\mathcal{N}=2$ hypermultiplets join a singlet to form the $\mathbf{16}$ (see~\cite{M05a}).  In the previous example, we made a choice to truncate the $\mathcal{N}=1$ left-chiral multiplets in the $\mathbf{\overline{16}}$ of $SO(10)$ (and their right-chiral conjugates in the $\mathbf{16}$).  

\subsection{Tensor multiplet couplings}\label{sec:tensor multiplet couplings}

When a MESGT with $n_{V}$ abelian vector multiplets is gauged, the symmetry group of the Lagrangian is broken to the gauge group $K\subset G$.  The $n_{V}+1$ vector fields decompose into $K$-reps 
\[ \mathbf{n_{V}+1}=\mbox{\textbf{adj}}(K)\oplus \mbox{\textbf{non-singlets}}(K)\oplus \mbox{\textbf{singlets}}(K). \]   
Such a gauging requires the non-singlet vector fields to be dualized to anti-symmetric tensor fields~\cite{GZ:99dec} satisfying a field equation that serves as a ``self-duality" constraint~\cite{selfdual} (thus keeping the number of degrees of freedom the same):
\begin{equation} B^{M}_{\mu\nu}=c^{M}_{N}\,\epsilon^{\;\;\;\rho\sigma\lambda}_{\mu\nu}\,\partial_{[\rho}B^{N}_{\sigma\lambda]}+\cdots, \label{selfdual} \end{equation}   
where $c^{M}_{N}$ has dimensions of inverse mass; square brackets denote anti-symmetric permutations; and ellipses denote terms involving other fields.  A tensor field does not require an abelian invariance to remain massless.  

We have already discussed the scalar sector of a pure 5D YMESGT.  When tensor multiplets are coupled the scalar manifold is again a real Riemannian space, but which cannot be decomposed globally as a product of ``vector" and ``tensor" parts.  We can, of course, identify an orthogonal frame of scalars at each point of the manifold: the vector multiplets are associated with the combination $h^{I}_{\tilde{x}}\phi^{\tilde{x}}$ at a given point, while the tensor multiplets are associated with the independent combination $h^{M}_{\tilde{x}}\phi^{\tilde{x}}$.  Similarly, the combination of fermions $h^{I}_{\tilde{p}}\lambda^{\tilde{p}\,i}$ are associated with vector multiplets, while $h^{M}_{\tilde{p}}\lambda^{\tilde{p}\,i}$ with tensor multiplets.  We will write $\phi^{\tilde{x}}$ and $\phi^{\tilde{m}}$ to denote the scalar partners of the vector and tensors, respectively, at any given point of the scalar manifold.  Similarly, we write $\lambda^{\tilde{p}\,i}$ and $\lambda^{\tilde{\ell}\,i}$ as the fermionic partners of the vector and tensor fields, respectively.  It is then implicitly understood that the meaning of this notation is given by the above discussion. 

When tensors are present, we will use indices $I,J,K$ for 5D vector fields and $M,N,P$ for 5D tensor fields.  Then $n_{T}$ tensor multiplets are
\[
\{B^{M}_{\hat{\mu}\hat{\nu}},\lambda^{\tilde{\ell}\,i},\phi^{\tilde{m}}\}.
\]
To be consistent with the gauge symmetry, the components of the $C$-tensor are constrained to be~\cite{GZ:99dec}:
\begin{equation}
\begin{split}
C_{IMN}=\frac{\sqrt{6}}{2}\Omega_{NP}\Lambda^{P}_{IM}\\
C_{MNP}=0,\;\;\;\;C_{MIJ}=0,\label{Btrs}
\end{split}
\end{equation}
where $\Omega_{NP}$ is antisymmetric and $\Lambda^{P}_{IM}$ are symplectic $K$-representation matrices appearing in the $K$-transformation of the tensor fields: $\delta_{\alpha} B^{M}_{\mu\nu}=\alpha^{I}\Lambda^{M}_{IN}B^{N}_{\mu\nu}.$
Furthermore, $C_{IJK}$ must be a rank-three symmetric $K$-invariant tensor.  Note: We are assuming a compact or non-semisimple gauge group $K$; see~\cite{Bergshoeff} for more general couplings where $C_{MIJ}\neq 0$. 

The new or modified terms in the bosonic 5D Lagrangian involving are~\cite{GZ:99dec}
\[ \begin{split}
&\mathcal{L}_{T}=-\frac{\hat{e}}{4}\stackrel{\circ}{a}_{MN}B^{M}_{\hat{\mu}\hat{\nu}}\,B^{N\,\hat{\mu}\hat{\nu}}
-\frac{\hat{e}}{2}\stackrel{\circ}{a}_{IM}\mathcal{F}^{I}_{\hat{\mu}\hat{\nu}}\,B^{M\,\hat{\mu}\hat{\nu}} \\
&+ \frac{1}{4g}\epsilon^{\hat{\mu}\hat{\nu}\hat{\rho}\hat{\sigma}\hat{\lambda}}\,\Omega_{MN}B^{M}_{\hat{\mu}\hat{\nu}}\,\nabla_{\hat{\rho}}B^{N}_{\hat{\sigma}\hat{\lambda}}\\ 
&-\frac{e}{2}g_{\tilde{x}\tilde{y}}\mathcal{D}_{\hat{\mu}}\phi^{\tilde{x}}
\mathcal{D}^{\hat{\mu}}\phi^{\tilde{y}}+ \frac{1}{2\sqrt{6}}C_{MNI}\, \epsilon^{\hat{\mu}\hat{\nu}\hat{\rho}\hat{\sigma}\hat{\lambda}}B^{M}_{\hat{\mu}\hat{\nu}}\,B^{N}_{\hat{\rho}\hat{\sigma}}\,A^{I}_{\hat{\lambda}},
\end{split} \]
where $\phi^{\tilde{x}}$ are scalars in both vector and tensor multiplets, and 
\[
\mathcal{D}_{\hat{\mu}}\phi^{\tilde{x}}=\partial_{\hat{\mu}}\phi^{\tilde{x}}+K^{\tilde{x}}_{I}A^{I}_{\hat{\mu}},
\]
with $K^{\tilde{x}}_{I}$ the Killing vectors on the vector/tensor scalar manifold.  
The 5D field equations for the $B^{M}_{\hat{\mu}\hat{\nu}}$ are (in terms of forms) 
\begin{equation}\begin{split} &^\star DB^{M}=g\Omega^{MN}\stackrel{\circ}{a}_{M\tilde{I}}\mathcal{H}^{\tilde{I}},\\
&\;\;\mbox{where}\;\;\;\;\mathcal{H}^{\tilde{I}}=\left( \begin{array}{c}
\mathcal{F}^{I}\\
B^{M}\\
\end{array}
\right).\end{split}\label{ELeqn}\end{equation}
The presence of non-trivially charged tensors also introduces a scalar potential $P^{(T)}$ that was not present in the case of pure YMESGTs.  The term in the Lagrangian is~\cite{GZ:99dec} 
\begin{equation}\begin{split}
&\mathcal{L}_{P^{(T)}}=-2g^{2}\hat{e}\,W^{\tilde{p}}W^{\tilde{p}},\\
&\;\;\mbox{where}\;\;\;\;W^{\tilde{p}}=-\frac{\sqrt{6}}{8}h^{\tilde{p}}_{M}\Omega^{MN}h_{N}.\end{split} \label{Tpot}
\end{equation}\vspace{1mm}\\
\textbf{Dimensional reduction}

In the dimensional reduction, we parametrize the tensor field as
\[ B^{M}_{\hat{\mu}\hat{\nu}}=
\left(
\begin{array}{ccc}
0 & & -\tilde{A}^{M}_{\nu}\\
\tilde{A}^{M}_{\mu} & & B^{M}_{\mu\nu} 
\end{array}
\right), \] 
where tildes have been used to help distinguish from 4D vectors arising as components of 5D vectors.  The resulting Lagrangian containing both $\tilde{A}^{M}_{\mu}$ and $B^{M}_{\mu\nu}$ is analogous to the 1st order formulation of the Freedman-Townsend model~\cite{Freedman}.  One can obtain a 2nd order formulation in terms of the $\tilde{A}^{M}_{\mu}$ by using the Euler-Lagrange equations, which appear as constraints relating these fields in the dimensionally reduced theory.  However, in the case of dimensional reduction, there is an obstruction~\cite{GMZ05b} to obtaining a \textit{local} 2nd order Lagrangian in terms of the $B^{M}_{\mu\nu}$.  This will not be present in the case of the orbifold.        

The $\xi^{5}$ transformations of the dimensionally reduced fields $B^{M}_{\mu\nu}$ and $\tilde{A}^{M}_{\mu}$ are
\begin{equation} \begin{split}
\delta_{\xi^{5}}B^{M}_{\mu\nu}&=\partial_{\mu}\xi^{5}\tilde{A}^{M}_{\nu}-\partial_{\nu}\xi^{5}\tilde{A}^{M}_{\mu}\\
\delta_{\xi^{5}}\tilde{A}^{M}_{\mu}&=0.
\end{split} \label{tensorKKtrs} \end{equation}
We must therefore make a field redefinition
\begin{equation}B^{M}_{\mu\nu}\rightarrow B^{M}_{\mu\nu}-4C_{[\mu}\tilde{A}^{M}_{\nu]}\label{BKKtrs}\end{equation}  
so that the $B^{M}_{\mu\nu}$ are now KK-invariant.  The full dimensional reduction of tensor-coupled YMESTs is given in~\cite{GMZ05b}.  The new or modified terms in the 5D bosonic Lagrangian are  
 \begin{equation} \begin{split}
 &e^{-1}\mathcal{L}^{(4)} = -e^{-2\sigma}{\stackrel{\circ}{a}}_{IM}(D_{\mu}A^{I})\tilde{A}^{\mu M}
-\frac{1}{2}e^{-2\sigma}{\stackrel{\circ}{a}}_{MN} \tilde{A}_{\mu}^{M}\tilde{A}^{\mu N}\\
&-\frac{3}{4}{\stackrel{\circ}{a}}_{\tilde{I}\tilde{J}}(\mathcal{D}_{\mu}\tilde{h}^{\tilde{I}})(\mathcal{D}^{\mu}\tilde{h}^{\tilde{J}})+\frac{e^{-1}}{2\sqrt{6}} C_{MNI}\epsilon^{\mu\nu\rho\sigma}B_{\mu\nu}^{M}
B_{\rho\sigma}^{N} A^{I}
\\
&+\frac{e^{-1}}{g} \epsilon^{\mu\nu\rho\sigma}\Omega_{MN}C_{\mu\nu}\tilde{A}_{\rho}^{M}\tilde{A}_{\sigma}^{N}  +\frac{e^{-1}}{g} \epsilon^{\mu\nu\rho\sigma}\Omega_{MN}B_{\mu\nu}^{M} D_{\rho}\tilde{A}^{N}_{\sigma}\\
&-\frac{1}{4}
e^{\sigma}{\stackrel{\circ}{a}}_{MN}B_{\mu\nu}^{M}B^{N\mu\nu}
-\frac{1}{2}e^{\sigma}{\stackrel{\circ}{a}}_{IM}(\mathcal{F}_{\mu\nu}^{I}+2C_{\mu\nu}A^{I})B^{M\mu\nu}\\
&-g^2 P,  \label{redlag2}
\end{split}\end{equation}
where
\begin{eqnarray}
D_{\mu}\tilde{A}^{M}_{\nu} & \equiv & \partial_{\mu} \tilde{A}_{\nu}^{N}+ g A_{\mu}^{I}\Lambda_{IP}^{N}\tilde{A}_{\nu}^{P}\\
\mathcal{D}_{\mu}\tilde{h}^{\tilde{I}}& \equiv & \partial_{\mu} \tilde{h}^{\tilde{I}} + g A_{\mu}^{I}M_{I\tilde{K}}^{\tilde{I}}\tilde{h}^{\tilde{K}}.
\end{eqnarray}
The total scalar potential, $P$, is now
\begin{equation}\begin{split}
P=&\;2e^{-\sigma}W^{\tilde{p}}W^{\tilde{p}}\\
&\;+\frac{3}{4} e^{-3\sigma}{\stackrel{\circ}{a}}_{\tilde{I}\tilde{J}} (A^{I}M_{I\tilde{K}}^{\tilde{I}} h^{\tilde{K}})
(A^{J}M_{J\tilde{L}}^{\tilde{J}}h^{\tilde{L}}), \label{totalpot}
\end{split}\end{equation}
where $W^{\tilde{p}}$ is defined in eqn~(\ref{Tpot}), and
\begin{equation}
     M_{(I)\tilde{K}}^{\tilde{J}}=   \left( \begin{array}{cc}
     f_{IK}^{J} & 0 \\
     0 & \Lambda_{IM}^{N} \end{array} \right), \label{MfB}
     \end{equation}
and $\tilde{I}=(I,M)$.\vspace{1mm}\\
\textbf{Parity assignments of tensor sector}

Since~(\ref{tensorKKtrs}) is only true for KK-transformations connected to the identity, the $\tilde{A}^{M}_{\mu}$ are not necessarily even under $\mathbb{Z}_{2}$ action.  However, the above do lead to the constraint
\[P(\tilde{A}^{M}_{\mu})=-P(B^{M}_{\mu\nu}),\] 
componentwise.  These two fields do not describe independent propagating degrees of freedom since they are related by a constraint equation (coming from the fact that the 5D tensors satisfied a ``self-duality" field equation~(\ref{selfdual}) reducing the number of propagating modes): 
\begin{equation}
B^{M}_{\mu\nu}=c^{M}_{N} (^{\star}D\tilde{A}^{N})_{\mu\nu}+\cdots,\label{sdtext}
\end{equation}
where $c^{M}_{N}$ is proportional to $\Omega^{MP} \stackrel{\circ}{a}_{PN}$; $\star$ is the Hodge operator; and the dots indicate terms involving other fields.  There are two classes of assignments we can make, characterized by the parity of the symplectic form $\Omega_{MN}$ on the vector space spanned by the 5D tensors.

\subsubsection{Odd Parity $\Omega_{MN}$}

If the self-duality relation is used to express all tensor fields in terms of the vectors $\tilde{A}^{M}_{\mu}$, the mass of the $\tilde{A}^{M}_{\mu}$ is non-vanishing at the orbifold fixed points.  However, there are insufficient fermionic degrees of freedom to form massive $\mathcal{N}=1$ vector multiplets.  Therefore, we must use the Euler-Lagrange equations to write $\tilde{A}^{M}_{\mu}\rightarrow B^{M}_{\mu\nu}$.  There is an obstruction to doing this with the dimensionally reduced 1st order Lagrangian obtained by using KK-invariant field redefinitions~\cite{GMZ05b}, in particular due to the topological term of the form $\Omega_{MN}C_{\mu\nu}\tilde{A}^{M}_{\mu}\tilde{A}^{N}_{\nu}$, but the problem is avoided in the \textit{orbifold} theory since the obstruction vanishes on the 4D fixed planes.   

Once the 2nd order Lagrangian with tensors is obtained, the $B^{M}_{\mu\nu}$ can then be Hodge dualized to scalars $B^{M}$ by adding a term of the form 
\begin{equation}
e^{-1}\Delta\mathcal{L}=\epsilon^{\mu\nu\rho\sigma}\Omega_{MN}\mathcal{B}^{M}_{\mu\nu\rho}\,D_{\sigma}B^{N}
\end{equation}
to the Lagrangian (if this step is performed before integrating over the fifth dimension, one needs a $\delta(x^{5})$ factor), where $D_{\rho}$ is the gauge covariant derivative acting on the scalars, and $\mathcal{B}^{M}_{\mu\nu\rho}=3!\,\partial_{[\mu}B^{M}_{\nu\rho]}$.   

The multiplets that will propagate on the fixed planes are
\begin{center}\begin{tabular}{|ccc|}\hline
Multiplet & Representation & Type\\\hline
$\{\bar{\lambda}^{\ell\,i},\;z^{M}\}$ & $\mathbf{N}\oplus\mathbf{\overline{N}}$ & $\mathbb{R}$\\\hline
\end{tabular}\end{center}
That is, there are left-chiral multiplets in a real representation along with the CPT conjugate right-chiral multiplets.\vspace{1mm}\\
\textbf{Example}

Consider the ``unified" 5D MESGT based on the Lorentzian Jordan algebra $J^{\mathbb{C}}_{(1,5)}$, whose global symmetry group is $SU(5,1)$~\cite{GZ03}.  We can couple this theory to hypermultiplets whose scalars parametrize the particular scalar manifold (\ref{mfd}).  If we gauge the common $SU(5)\times U(1)\subset SU(5,1)$ subgroup, we will get $SU(5)\times U(1)$ gauge multiplets, along with tensor multiplets in the $\mathbf{5}\oplus\mathbf{\bar{5}}$ and hypermultiplets in the $\mathbf{1}\oplus\mathbf{5}\oplus\mathbf{10}$.  This is then similar to the ground state theory in the $SU(5,1)$ gauging example before, but with some important differences; one of which being that the 5D bare graviphoton $A^{0}_{\hat{\mu}}$ does not take part in gauging the isometries of the scalar manifold.  We can make the assignments
\begin{center}\begin{tabular}{|c|c|c|c|c|c|}\hline
$\alpha$ & $a$ & $0$ & $M$ & $n$ & $\tilde{n}$\\\hline
$\mbox{adj}[SM\times U(1)]$ & $SU(5)/SM$ & $\mathbf{1}$ & $\mathbf{5}\oplus\mathbf{\bar{5}}$ & $\mathbf{1}\oplus\mathbf{\bar{5}}\oplus\mathbf{10}$ & $c.c.$\\\hline
\end{tabular} \end{center}
The propagating modes along the fixed planes will be $SU(3)\times SU(2)\times U(1)^{2}$ gauge fields; weak doublet (Higgs) chiral multiplets; color triplet chiral multiplets; and left-chiral matter multiplets (including a sterile fermion multiplet) with their CPT conjugates.  Again, there is also the generic singlet spin-1/2 multiplet coming from the 5D supergravity multiplet, and chiral multiplets in the symmetric space $SU(5)/SM$.  All of these multiplets are tree-level massless, while the scalars in the $\mathbf{5}\oplus\mathbf{\bar{5}}$ have a potential term.  

\subsubsection{Even Parity for $\Omega_{MN}$}

If $\Omega_{MN}$ has even parity, then the $\mathbb{Z}_{2}$ action acts reducibly on the symplectic vector space and projects out half of the bosonic fields.  Consistency with supersymmetry requires that we use the field equation (\ref{sdtext}) to eliminate $B^{M}_{\mu\nu}$ from the Lagrangian.  Splitting $M=(\mathcal{M},\bar{\mathcal{M}})$ with $\mathcal{M}=1,\ldots,n_{T}/2$ and $\bar{\mathcal{M}}=n_{T}/2,\ldots, n_{T}$, the $\tilde{A}^{\mathcal{M}}_{\mu}$ then have a mass matrix $$M_{\mathcal{M}\mathcal{N}}\equiv e^{-2\sigma}\stackrel{\circ}{a}_{\mathcal{M}\mathcal{N}}.$$  Therefore, the $n_{T}/2$ vectors, $n_{T}$ spin-1/2 fields, and $n_{T}/2$ scalars form $n_{T}/2$ massive $\mathcal{N}=1$ vector multiplets.  

The multiplets with propagating modes on the boundaries are
\begin{center}\begin{tabular}{|ccc|}\hline
Multiplet & Representation & Type\\\hline
$\{A^{\mathcal{M}},\;\lambda^{\bar{\ell}\,i},\;h^{\mathcal{M}}\}$ & $\mathbf{N}_{\mathbb{R}}$ & $\mathbb{R}$\\\hline
\end{tabular}\end{center}
The notation for the representation means that the gauge group at the fixed points must support a \textit{real} $\mathbf{N}$ (whereas the 5D gauge group had to support a complex $\mathbf{N}$).  Let's illustrate this with an example.\vspace{1mm}\\
\textbf{Example}

The minimal example in which one is left with a group containing SM is where the 5D gauge group is $SU(10)\times U(1)$.  Starting with the ``unified" MESGT defined by the Lorentzian Jordan algebra $J^{\mathbb{C}}_{(1,10)}$ and with $SU(N,1)$ global symmetry of the Lagrangian, we can gauge the $SU(10)\times U(1)$ subgroup, yielding tensors in the $\mathbf{10}\oplus\mathbf{\overline{10}}$.  If the symplectic form has even parity, then the orbifold conditions require the group to be broken to at least $SO(10)\times U(1)$, under which we have massive vector multiplets in the (real) $\mathbf{10}$.  This theory will appear as a ground state of the 4D $\mathcal{N}=1$ theory with gauge group $(SO(10)\times U(1))\ltimes T^{10}$, analogous to the $\mathcal{N}=2$ case discussed in~\cite{GMZ05b}.  There are also chiral multiplets from the broken gauge multiplets forming the $\mathbf{54}$, along with their CPT conjugates.

\section{Objects other than fields}\label{sec:objects other than fields}

There are field-dependent and independent objects that appear in the Lagrangian and supersymmetry transformations that carry $\mathbb{Z}_{2}$ parity.  In particular, the field independent objects are the $C_{IJK}$ tensor defining the MESGT that exists prior to gauging; the structure constants $f^{I}_{JK}$ and transformation parameters $\alpha^{I}(x)$ of the 5D gauge group; and the symplectic tensor $\Omega_{MN}$ and transformation matrices $\Lambda^{M}_{IN}$ in the tensor coupled theory.  \textit{These contain an implicit $\epsilon(x^{5})$ or $\kappa(x^{5})$ factor when assigned odd parity}.  The field dependent objects are the restricted ambient space metric $\stackrel{\circ}{a}_{IJ}(\phi)$ and scalar manifold metrics $g_{xy}(\phi)$ and $g_{XY}(q)$; the $h^{I}_{p}(\phi)$; the scalar vielbein $f^{p}_{x}(\phi)$ and $f^{X}_{i\,A}(q)$; the Killing vectors on the scalar manifold $K^{I}_{x}(\phi)$ and $K^{I}_{X}(q)$.  \textit{These vanish when assigned odd parity} due to the form of the bosonic field expansion in eqn~(\ref{phi}).\vspace{1mm}\\
\textbf{Pure YMESGT}

In~\cite{MESGT}, it was shown that the $C_{IJK}$ defining a MESGT may be put in a ``canonical" basis satisfying the positivity of $\mathcal{V}=C_{IJK}\zeta^{I}\zeta^{J}\zeta^{K}$ (see~(\ref{canonical})).
The parity assignments of the components are determined by requiring the polynomial $\mathcal{V}$ to be invariant under $\mathbb{Z}_{2}$ action.  Splitting $i=(\alpha, a)$, we have
\begin{center}\begin{tabular}{|c|c|}  \hline
Even & Odd \\ \hline  
$C_{000}\;\;C_{0ab}\;\;C_{0\alpha\beta}\;\;C_{abc}\;\;C_{a\alpha\beta}$ & $C_{0a\alpha}\;\;C_{\alpha\beta\gamma}\;\;C_{\alpha ab}$ \\\hline
\end{tabular} \end{center}
\textit{There is freedom in choosing $\epsilon(x^{5})$ or $\kappa(x^{5})$ as the jumping function for odd components.} 

Consistency of the infinitesimal gauge transformations~(\ref{alpha}) require parity assignments for the $f^{I}_{JK}$ and $\alpha^{I}(x)$ to be 
\begin{center}\begin{tabular}{|c|c|}  \hline
Even & Odd \\ \hline  
$f^{\alpha}_{\beta\gamma}\;\;f^{\alpha}_{ab}$ & $f^{a}_{bc}\;\;f^{a}_{\alpha\beta}$\\
$f^{0}_{a \beta}\;\;f^{0}_{\alpha 0}$ & $f^{0}_{\alpha\beta}$\\
$\alpha^{\beta}$ & $\alpha^{0}\;\;\alpha^{b}$\\ \hline
\end{tabular} \end{center}
where $f^{I}_{JK}$ vanishes if any of the indices correspond to 5D spectator vector fields~\footnote{This will be true, e.g., for the ``bare graviphoton" $A^{0}_{\mu}$ if the 5D gauge group is compact.}; and permutations of the indices have the same parity.
The (upstairs picture) gauge transformation parameters are subject to an expansion on $S^{1}/\mathbb{Z}_{2}$ just as in (\ref{phi}).  Consistency of the 5D gauge algebra requires that \textit{odd $f^{I}_{JK}$ be redefined by $\epsilon(x^{5})f^{I}_{JK}$}, where the $f^{I}_{JK}$ are now even~\footnote{Note that the orbifold fixed planes are not oriented surfaces.}.      

The components of the restricted ambient space metric and scalar manifold metric have parities determined by the requirement that the line elements of those spaces be preserved:  
\begin{center}\begin{tabular}{|c|c|}  \hline
Even & Odd \\ \hline  
$\stackrel{\circ}{a}_{\alpha\beta}\;\;\stackrel{\circ}{a}_{ab}$ & $\stackrel{\circ}{a}_{a\beta}\;\;\stackrel{\circ}{a}_{0\alpha}$\\
$\stackrel{\circ}{a}_{00}\;\;\stackrel{\circ}{a}_{0a}$ & $$\\
$g_{xy}\;\;g_{\chi\psi}$ & $g_{\chi y}$ \\\hline
\end{tabular} \end{center}
Consistency of the gauge transformations~(\ref{alpha}) determine the parities of the scalar manifold Killing fields:
\begin{center}\begin{tabular}{|c|c|}  \hline
Even & Odd \\ \hline  
$K^{x}_{\alpha}$ & $K^{x}_{0}\;\;K^{x}_{a}$\\
$K^{\chi}_{0}\;\;K^{\chi}_{a}$ & $K^{\chi}_{\alpha}$
\\ \hline 
\end{tabular}\end{center}
The non-zero components $K^{\chi}_{a}$, which are a set of sections of the normal bundle over the 4D scalar manifold, determine the form of the scalar potential involving the $\phi^{x}$ and $A^{a}$ at the fixed points:
\begin{equation}
g_{\chi\psi}\,K^{\chi}_{a}\,K^{\psi}_{b}A^{a}A^{b},\label{potA}
\end{equation}
where $g_{\chi\psi}$ is the metric determined by the normal bundle connection.

Finally, the functions $h^{I}$ and $h^{I}_{p}$; the scalar vielbein $f^{p}_{x}$; and the functions $h^{I}_{x}=h^{I}_{p}f^{p}_{x}$ are required to satisfy
\begin{center}\begin{tabular}{|c|c|}  \hline
Even & Odd \\ \hline  
$h^{0}\;\;h^{a}$ & $h^{\alpha}$\\
$h^{0}_{p}\;\;h^{a}_{p}\;\;h^{\alpha}_{\rho}$ & $h^{a}_{\rho}\;\;h^{\alpha}_{p}$\\
$f^{p}_{x}\;\;f^{\rho}_{\chi}$ & $f^{p}_{\chi}\;\;f^{\rho}_{x}$\\
$h^{0}_{x}\;\;h^{a}_{x}\;\;h^{\alpha}_{\chi}$ & $h^{a}_{\chi}\;\;h^{\alpha}_{x}$\\ \hline
\end{tabular}  \end{center}
\textbf{Tensor couplings}

In the tensor-coupled theory, parity assignments depend on the parity of the symplectic form $\Omega_{MN}$.  The parities of the additional $C$-tensor components must be  
\begin{center}\begin{tabular}{|c|c|} \hline
Even & Odd \\ \hline
$C_{\mathcal{M}\bar{\mathcal{N}}\alpha}$ & $C_{\mathcal{M}\mathcal{N}\alpha}\;\;C_{\bar{\mathcal{M}}\bar{\mathcal{N}}\alpha}$ \\ 
$C_{\mathcal{M}\mathcal{N}a}\;\;C_{\bar{\mathcal{M}}\bar{\mathcal{N}}a}$ & $C_{\mathcal{M}\bar{\mathcal{N}}a}$ \\ \hline 
\multicolumn{2}{c}{$P(\Omega_{MN})=+\Omega_{MN}$}\end{tabular}\;\;\;\;\;\;\;\;\;\;\begin{tabular}{|c|c|}\hline
Even & Odd\\\hline
$C_{MNa}$ & $C_{MN\alpha}$ \\ \hline 
\multicolumn{2}{c}{$P(\Omega_{MN})=-\Omega_{MN}$}
\end{tabular} \end{center}
Consistency of the gauge transformations require the representation matrices to satisfy
\begin{center}\begin{tabular}{|c|c|} \hline 
Even & Odd \\\hline
$\Lambda^{\mathcal{M}}_{\alpha \,\mathcal{N}}\;\;\Lambda^{\bar{\mathcal{M}}}_{\alpha\,\bar{\mathcal{N}}}$ & $\Lambda^{\bar{\mathcal{M}}}_{\alpha \,\mathcal{N}}\;\;\Lambda^{\mathcal{M}}_{\alpha\,\bar{\mathcal{N}}}$\\
$\Lambda^{\bar{\mathcal{M}}}_{0 \,\mathcal{N}}\;\;\Lambda^{\mathcal{M}}_{0\,\bar{\mathcal{N}}}$ & $\Lambda^{\bar{\mathcal{M}}}_{0\,\bar{\mathcal{N}}}\;\;\Lambda^{\mathcal{M}}_{0 \,\mathcal{N}}$\\
$\Lambda^{\bar{\mathcal{M}}}_{a \,\mathcal{N}}\;\;\Lambda^{\mathcal{M}}_{a\,\bar{\mathcal{N}}}$ & $\Lambda^{\bar{\mathcal{M}}}_{a\,\bar{\mathcal{N}}}\;\;\Lambda^{\mathcal{M}}_{a \,\mathcal{N}}$\\\hline 
\multicolumn{2}{c}{$P(\Omega_{MN})=+\Omega_{MN}$}  
\end{tabular}\;\;\;\;\;\;\;\;\;\;
\begin{tabular}{|c|c|} \hline
Even & Odd \\\hline
$\Lambda^{M}_{\alpha N}$ & $\Lambda^{M}_{0 N}\;\;\Lambda^{M}_{a N}$\\ \hline 
\multicolumn{2}{c}{$P(\Omega_{MN})=-\Omega_{MN}$}
\end{tabular}\end{center}
If odd gauge transformation parameters $\alpha^{a}$ are expanded as in eqn~(\ref{phi}), odd $\Lambda^{M}_{IN}\rightarrow \epsilon(x^{5})\Lambda^{M}_{IN}$ for consistency of the gauge algebra; the relation $C_{MNI}\propto \Omega_{MP}\Lambda^{P}_{IN}$ from eqn (\ref{Btrs}) then requires odd $C_{IMN}\rightarrow \epsilon(x^{5})C_{IMN}$.  Note that when $\Omega_{MN}$ has odd parity, the set of coefficients $c^{M}_{N}\sim \Omega^{MP}\stackrel{\circ}{a}_{PN}$ are odd; a choice that was made in~\cite{YL:03dec}.  

As in the pure YMESGT case, the ambient space and scalar manifold line elements should be preserved under the $\mathbb{Z}_{2}$ action so that 
\begin{center}\begin{tabular}{|c|c|} \hline 
Even & Odd \\\hline
$\stackrel{\circ}{a}_{\mathcal{M}\mathcal{N}}\;\;\stackrel{\circ}{a}_{\bar{\mathcal{M}}\bar{\mathcal{N}}}$ & $\stackrel{\circ}{a}_{\mathcal{M}\bar{\mathcal{N}}}$\\
$\stackrel{\circ}{a}_{\bar{\mathcal{M}}\alpha}\;\;\stackrel{\circ}{a}_{\mathcal{M} a}$ & $\stackrel{\circ}{a}_{\mathcal{M}\alpha}\;\;\stackrel{\circ}{a}_{\bar{\mathcal{M}}a}$\\
$g_{\tilde{x}\tilde{y}}\;\;g_{mn}$ & $g_{\tilde{x}m}$ \\\hline
\multicolumn{2}{c}{$P(\Omega_{MN})=+\Omega_{MN}$}
\end{tabular}\;\;\;\;\;\;\;\;\;\;
\begin{tabular}{|c|c|} \hline
Even & Odd \\\hline
$\stackrel{\circ}{a}_{aM}$ & $\stackrel{\circ}{a}_{\alpha M}$\\
$g_{\tilde{x}\tilde{y}}\;\;g_{mn}$ & $g_{\tilde{x}m}$ \\\hline
\multicolumn{2}{c}{$P(\Omega_{MN})=-\Omega_{MN}$} 
\end{tabular}\end{center}
Finally, the functions $h^{M}(\phi)$ and $h^{M}_{\ell}$; the vielbein $f^{\ell}_{m}$; and the functions $h^{I}_{x}=f^{\ell}_{x}h^{I}_{\ell}$ satisfy
 \begin{center}\begin{tabular}{|c|c|}  \hline
Even & Odd \\ \hline 
$h^{\mathcal{M}}(\phi)$& $h^{\bar{\mathcal{M}}}(\phi)$ \\ 
$h^{\mathcal{M}}_{m}\;\;h^{\bar{\mathcal{M}}}_{\bar{m}}$ & $h^{\mathcal{M}}_{\bar{m}}\;\;h^{\bar{\mathcal{M}}}_{m}$\\
$h^{\mathcal{M}}_{\ell}\;\;h^{\tilde{\mathcal{M}}}_{\bar{\ell}}$ & $h^{\tilde{\mathcal{M}}}_{\ell}\;\;h^{\mathcal{M}}_{\bar{\ell}}$\\\hline
\multicolumn{2}{c}{$P(\Omega_{MN})=+\Omega_{MN}$}
\end{tabular}\;\;\;\;\;\;\;\;\;\;
\begin{tabular}{|c|c|}  \hline
Even & Odd \\ \hline 
$h^{M}(\phi)$ &  \\
$h^{M}_{m}$ & $$ \\
$h^{M}_{\ell}$ & $h^{M}_{\bar{\ell}}$\\\hline
\multicolumn{2}{c}{$P(\Omega_{MN})=-\Omega_{MN}$}
\end{tabular}\end{center}
where we have split $\tilde{\ell}=(\ell, \bar{\ell})$ and $\tilde{m}=(m,\bar{m})$.\vspace{1mm}\\
\textbf{Hypermultiplet couplings}      

The parity assigments for the Killing vectors and vielbein of the quaternionic scalar manifold are required to be 
\begin{center}\begin{tabular}{|c|c|}  \hline
Even & Odd \\ \hline 
$K^{X_{1}}_{\alpha}\;\;K^{X_{2}}_{a}$ & $K^{X_{1}}_{a}\;\;K^{X_{2}}_{\alpha}$ \\ 
$f^{X_{1}}_{1A_{1}}\;\;f^{X_{1}}_{2A_{2}}$ & $f^{X_{1}}_{2A_{1}}\;\;f^{X_{1}}_{1A_{2}}$\\
$f^{X_{2}}_{2A_{1}}\;\;f^{X_{2}}_{1A_{2}}$ & $f^{X_{2}}_{1A_{1}}\;\;f^{X_{2}}_{2A_{2}}$\\\hline 
\end{tabular} \end{center}

\section{Parity assignments of the scalar sector}\label{sec:constraints on assigning parities}

In the previous sections, we listed the boundary multiplets by ``orbifolding" 5D YMESGTs coupled to tensor multiplets.  It's clear, for example, that 5D vector multiplets should yield 4D boundary vector multiplets when the associated scalar manifold embedding functions $h^{I}(\phi)$ are assigned odd parity, so that the parities of the $h^{I}$ induce parity assignments for the physical scalars $\phi^{\tilde{x}}$.  However, it is not immediately clear \textit{how} the assignments are transmitted from one to the other.  That is, how are the physical scalars truncated in light of the fact that the `odd' scalar partners of the `even' vectors $A^{\alpha}_{\hat{\mu}}$ are the combinations $\vev{h^{\alpha}_{\tilde{x}}}\phi^{\tilde{x}}$, where $h^{\alpha}_{\tilde{x}}\propto h^{\alpha}_{,\,\tilde{x}}$.  We will consider this using two examples based on symmetric ``very special" real scalar manifolds.        

The first example is of the ``generic Jordan" family of Maxwell-Einstein supergravity theories (MESGTs) with symmetric scalar manifolds of the form~\cite{MESGT} 
\[
\mathcal{M}_{R}=SO(1,1)\times \frac{SO(n_{V}-1,1)}{SO(n_{V}-1)},
\]
where $n_{V}$ is the number of vector multiplets.  
The cubic polynomial for the theory in the absence of an orbifold is $\mathcal{V}=C_{IJK}
\xi^{I}\xi^{J}\xi^{K}$, where 
\begin{gather}
C_{000}=1,\;\;C_{00i}=0,\;\;C_{0ij}=-\frac{1}{2}\delta_{ij},\nonumber\\
C_{111}=\frac{1}{\sqrt{2}},\;\;C_{1ab}=+\frac{1}{\sqrt{2}}\delta_{ab},\nonumber
\end{gather}
with $a,b=2,\ldots,n_{V}$.
On $\mathcal{M}_{4}\times S^{1}/\mathbb{Z}_{2}$, the $C_{IJK}$ can have odd components satisfying jumping conditions, though we take $h^{0}$ to always have even parity.  

For example, assigning odd parity to $h^{1}$ and $C_{1ij}$ (and redefining $C_{1ij}\rightarrow \epsilon(x^{5})C_{1ij}$), the polynomial is 
\[\begin{split}
\mathcal{V}=&\,\left(\frac{2}{3}\right)^{\frac{3}{2}} \left[ (\xi^{0})^{3}-\frac{3}{2}\xi^{0}\delta_{ij}\xi^{i}\xi^{j}-\frac{\epsilon(x^{5})}{\sqrt{2}}(\xi^{1})^{3}\right. \\
&\;\;\;\;\;\;\;\;\;\;\;\;\;\;\;\left. +\frac{3\epsilon(x^{5})}{\sqrt{2}}\xi^{1}[(\xi^{2})^{2}+\cdots+(\xi^{n_{V}})^{2}] \right] .\end{split}
\]
Restricting to the scalar hypersurface $\mathcal{V}=1$, the terms with $h^{1}\equiv \xi^{1}|_{\mathcal{V}=1}$ vanish (since $h^{1}$ is a parity-odd function of scalars) so that, at the orbifold fixed points,
\[\mathcal{V}|_{fp}=h^{0}\left\{(h^{0})^{2}-\frac{3}{2}\delta_{ab}h^{a}h^{b}\right\}.\] 
\textbf{Remark:}  In general, 4D $\mathcal{N}=1$ supergravity theories are in 1-1 correspondence with Hodge manifolds.  The 4D $\mathcal{N}=1$ supergravity theory we obtain from orbifolding is of a special class based on a (not necessarily irreducible) cubic polynomial satisfying $\mathcal{V}_{fp}(\mbox{Im} (z))=e^{3\sigma}>0$.

In the family of `generic Jordan' MESGTs, however, the solution to the condition $\mathcal{V}=1$ is simplest by making linear transformations $\check{\xi}^{I}=M^{I}_{J} \xi^{J}$ to a `non-canonical' basis~\cite{scalarmfds} (before orbifolding the theory):
\[\begin{split}
&M^{0}_{0}=\frac{1}{\sqrt{3}},\;\;\;\;M^{0}_{1}=-\sqrt{\frac{2}{3}}\\
&M^{1}_{0}=\sqrt{\frac{2}{3}},\;\;\;\;M^{1}_{1}=\frac{1}{\sqrt{3}}\\
&M^{\jmath}_{\imath}=\delta^{\jmath}_{\imath},
\end{split}\]
where $\imath,\jmath=2,\ldots,n_{V}$; and all other $M^{I}_{J}$ are zero.  The solution is then
 \[\begin{split}
&\mathcal{V}=\frac{3^{3/2}}{2}\check{h}^0 ||\check{h}||^{2}=1\\
&\check{h}^{0}= \frac{1}{\sqrt{3}||\phi||^2},\;\;\;\;\check{h}^{i}= \sqrt{\frac{2}{3}}\phi^{i},
\end{split}
\] 
where $i=1,\ldots, n_{V}$ and $||A||$ denotes the Minkowski norm with signature $(+-\cdots -)$.  In this basis, assigning odd parity to the $h^{\alpha}$ is equivalent to assigning odd parity to the $\phi^{\alpha}$, and truncation of the scalar $\vev{h^{\alpha}_{\tilde{x}}}\phi^{\tilde{x}}$ is obtained by $\vev{\phi^{\alpha}}=0$.   
  
Let's contrast this with another example: the ``generic non-Jordan" family~\cite{MESGT} of MESGTs with cubic polynomial
\[
\mathcal{V}=\sqrt{2}\check{\xi}^{0}(\check{\xi}^{1})^{2}-\check{\xi}^{1}\sum_{\jmath} (\check{\xi}^{\,\jmath})^{2},
\]
with solution to $\mathcal{V}=1$~\cite{scalarmfds}:
\[ \begin{split}
\check{h}^{0}=\,&\frac{1}{\sqrt{3}(\phi^{1})^{2}}+\frac{1}{\sqrt{3}}\phi^{1}\sum_{\jmath}(\phi^{\,\jmath})^{2}\\
\check{h}^{1}=\,&\left(2/3\right)^{\frac{3}{2}}\phi^{1}\\
\check{h}^{\jmath}=\,&\left(2/3\right)^{\frac{3}{2}}\phi^{1}\phi^{\,\jmath}
\end{split} \]
where $\jmath=2,\ldots, n_{V}$ and $\check{h}^{I}$ is not in the canonical basis.\\
\textbf{Remarks}\\
(i) The above solution requires that $\check{h}^{0}$ and $\check{h}^{1}$ must have the same parity. \\
(ii) The truncation of scalars $h^{\alpha}_{\tilde{x}}\phi^{\tilde{x}}$ still corresponds to the truncation $\phi^{\alpha}=0.$\\
The first remark tells us that assigning parities consistently in a given YMESGT requires a choice of basis $\xi^{I}$, and the solution to the condition $\mathcal{V}=1$.  As for the second remark, the requirement that the scalar combination $\vev{\check{h}^{\alpha}_{\tilde{x}}}\phi^{\tilde{x}}$ be odd, and so vanish at the orbifold fixed points, allows for a continuous family of vacua parametrized by the scalar vevs.  For example, if $A^{2}_{\mu}$ has even parity, then $\check{h}^{2}_{\tilde{x}}\phi^{\tilde{x}}$ must have odd parity so that there is a family of vacua with the direction normal to the flows being $\vev{\phi^{1}}\partial_{\phi^{2}}+\vev{\phi^{2}}\partial_{\phi^{1}}$ (this is the direction in which the propagating scalar is truncated).  However, as $\check{h}^{2}$ must also be odd, this requires $\phi^{2}$ to be odd so that it vanishes at the orbifold fixed points.  The point is that the theory is simply restricted to $\vev{\phi^{2}}=0$ vacua, connected to the basepoint $\check{h}^{I}=(1,0,\ldots,0)$ of the original 5D theory.  

We now turn briefly to the implications of odd parity assignments for the $h^{I}$.  Once we assign odd parities to a set $\xi^{\alpha}$, these must vanish on the fixed planes in a basis independent way (as well as $h^{\alpha}\propto \xi^{\alpha}|_{\mathcal{V}=1}$).  For this to be true, the theory must lie on the boundary of the classical K\"{a}hler cone, so that a consistent formulation of supergravity requires additional massless states located at the fixed points.  These, in turn, admit a higher dimensional interpretation as 11D supergravity~\cite{CYcomp} membranes wrapping collapsing Calabi-Yau 2-cycles~\cite{W96b}.  The construction of consistent minimal ($\mathcal{N}=2$) five-dimensional supergravity on the extended K\"{a}hler cone was studied in~\cite{mohaupt}; but in the present case, the theory on the fixed planes is 4D $\mathcal{N}=1$, which is a less restrictive framework to work with.  In addition, the quantum K\"{a}hler cone is generally different from the classical one~\cite{AGM}.  Therefore, the determination of these additional massless states is outside the scope of the present paper.  We end by noting that, due to the collapsing 2-cycles, the dual CY 4-cycles may collapse to 2- or 0-cycles; this is encoded in the functions $h_{I}=C_{IJK}h^{J}h^{K}$, which are rescaled 4-cycle volumes.  Therefore, the choice of jumping function for odd $C_{IJK}$ determines whether the 4-cycle collapses completely, and one must choose appropriately.             

\section{Extension to $\Gamma=\mathbb{Z}_{2}\times \mathbb{Z}_{2}$}\label{sec:extension to Gamma=Z2xZ2}

There are a couple of phenomenological issues that make the $S^{1}/\mathbb{Z}_{2}$ orbifold models too simplistic.  First, there are always massless chiral multiplets in real representations when a gauge group is broken at the orbifold fixed planes (though these may contain MSSM Higgs fields).  Second, all chiral multiplets come in complete representations of the 5D gauge group, which can lead to unwanted fields charged under the Standard Model gauge group.  The boundary conditions described by an $S^{1}/(\mathbb{Z}_{2}\times\mathbb{Z}_{2})$~\cite{ph/0011311} construction are for the most part capable of resolving these issues.  

An exception is the tensor sector: although there is a choice in assignment of parity for the symplectic form $\Omega_{MN}$, we cannot assign $(+-)$ parity under $\mathbb{Z}_{2}\times\mathbb{Z}_{2}$ action (it leads to inconsistencies in assignments for the fields).  Furthermore, given a choice of $\Omega_{MN}$ parity, there wasn't a choice of parity assignments in the $\Gamma=\mathbb{Z}_{2}$ case since supersymmetry dictated the results.  Therefore, the situation with tensors is no different in the $S^{1}/(\mathbb{Z}_{2}\times\mathbb{Z}_{2})$ construction.  \textit{This means that, e.g., tensor multiplets do not allow a doublet-triplet resolution via parity assignments} (see the example in section~\ref{sec:supermultiplets appearing in Gamma=Z2 case}). 

An expansion of a field $\Phi(x,x^{5})$ on $S^{1}/(\mathbb{Z}_{2}\times\mathbb{Z}_{2})$ will be of the form
\[ \begin{split}
\Phi^{(++)}(x,x^{5})&=\sum_{n} \Phi^{(n)}(x) \cos\left[2nx^{5}/R\right] \\
\Phi^{(+-)}(x,x^{5})&=\sum_{n} \Phi^{(n)}(x) (A_{(n)}\cos\left[(2n+1)x^{5}/R\right]+\cdots\\
\Phi^{(-+)}(x,x^{5})&=\sum_{n} \Phi^{(n)}(x) (C_{(n)}\sin\left[(2n+1)x^{5}/R\right]+\cdots\\  
\Phi^{(--)}(x,x^{5})&=\sum_{n} \Phi^{(n)}(x) (E_{(n)}\sin\left[2nx^{5}/R\right]+\cdots, 
\end{split}\]
where ellipses denote terms with $\epsilon(x^{5})$ factors as in eqn~(\ref{phi}), with expansion constants $B_{(n)}, D_{(n)}, F_{(n)}$, respectively.  Once again, bosonic fields cannot have $\epsilon$ factors since the upstairs picture Lagrangian and equations of motion would involve $\delta^{2}$, where $\delta$ is the Dirac distribution.  For those, we must set $B_{(n)}=D_{(n)}=F_{(n)}=0$.  Therefore, bosonic $\Phi^{(+-)}(x,x^{5})$ vanish at $x^{5}=0$ and bosonic $\Phi^{(-+)}(x,x^{5})$ vanish at $x^{5}=\pi R/2$; fermionic fields are not generally well-defined on the orbifold fixed planes.  Let $P(\Phi)$ be the parity of $\Phi$ under the first $\mathbb{Z}_{2}$ factor, and $P'(\Phi)$ denote the parity under the second factor.  Taking the covering space to be $[-\pi R,\pi R]$ (with $\{-\pi R\}\equiv \{\pi R\}$) as before, the orbifold now has fixed points at $\{0\},\{\pi R/2\}$.   

\subsection{Vector sector}\label{sec:vector sector 1}

In the previous sections, we made an index split for quantities with $\pm 1$ parity under the single $\mathbb{Z}_{2}$.  We will make a further index splitting for quantities with the four possible values $\{\pm 1,\pm 1\}$ for the parity $\{P(\Phi),P'(\Phi)\}$:
\[ i=(\alpha,\alpha',a,a')\;\;\;\;\tilde{p}=(\rho,\rho',p,p'). \]      
\textit{A given assignment of $\mathbb{Z}_{2}\times\mathbb{Z}_{2}$ parity to an object will consist of the union of two assignments in the $S^{1}/\mathbb{Z}_{2}$ construction}.  Fields from the 5D vector multiplets will have the following assignments:
\begin{center}\begin{tabular}{|cccc|}\hline
$++$ & $+-$ & $-+$ & $--$ \\ \hline
$A^{\alpha}_{\mu}$ & $A^{\alpha'}_{\mu}$ & $A^{a}_{\mu}$ & $A^{a'}_{\mu}$\\
$A^{a'}$ & $A^{a}$ & $A^{\alpha'}$ & $A^{\alpha}$\\
$h^{a'}$ & $h^{a}$ & $h^{\alpha'}$ & $h^{\alpha}$\\
$\delta^{\rho}$ & $\delta^{\rho'}$ & $\delta^{p}$ & $\delta^{p'}$\\
$\gamma^{p'}$ & $\gamma^{p}$ & $\gamma^{\rho'}$ & $\gamma^{\rho}$\\ \hline
\end{tabular} \end{center}
Note: the bare graviphoton $A^{0}_{\mu}$ always has $(--)$ parity (so $A^{0}$ has (++) parity).  The range of $\wp_{1}$, $\wp_{2}$, and $\wp_{3}$ in $\alpha=1,\ldots,\wp_{1}$; $\alpha'=\wp_{1}+1,\ldots, \wp_{2}$; $a=\wp_{2}+1,\ldots, \wp_{3}$; and $a'=\wp_{3}+1,\ldots, n_{V}$, are arbitrary.  

The fields with $(+-)$ or $(-+)$ eigenvalues have massive $n=0$ modes on the fixed planes for the same reason that any Kaluza-Klein field does: there is excitation in the $x^{5}$ direction.  In the low energy effective theory, such fields will fall into massive $\mathcal{N}=1$ multiplets in four dimensions due to terms in the Lagrangian with $\partial_{5}\Phi^{+-}$ or $\partial_{5}\Phi^{-+}$.    
  
In contrast to the $S^{1}/\mathbb{Z}_{2}$ construction, we can now choose to make massive multiplets out of unwanted light chiral multiplets in real representations by choosing there to be no $a',\,p'$ indices.  Alternatively, we can keep a subset of those massless chiral multiplets (in a real representation) \textit{such that they no longer furnish complete $K$-representations}.  The multiplets with propagating modes on the boundaries and their properties are listed below (we have decomposed the representation $R_{V}[K]=\mbox{adj}[K_{\alpha}]\oplus R^{1}_{V}[K_{\alpha}]\oplus R^{2}_{V}[K_{\alpha}]\oplus R^{3}_{V}[K_{\alpha}]$):
\begin{center}\begin{tabular}{|ccccc|}\hline
Multiplet & Representation & Type & Boundary & Mass\\\hline
$\{A^{\alpha}_{\mu},\lambda^{\rho\,i}\}$ & $\mbox{adj}[K_{\alpha}]$ & Real & Both & Massless\\
$\{\bar{\lambda}^{p'\,i},z^{a'}\}$ & $R^{1}_{V}[K_{\alpha}]$ & Real & Both & Potential\\
$\{\Psi^{i}_{5},z^{0}\}$ & $K_{\alpha}$-singlet & Real & Both & Massless\\
$\{A^{\alpha'}_{\mu}, \lambda^{\rho'\,i}\}$ & $R^{2}_{V}[K_{\alpha}]$ & Real & $y=0$ & $\mathcal{O}(1/R)$\\
$\{\bar{\lambda}^{p\,i}, z^{a}\}$ & $R^{3}_{V}[K_{\alpha}]$ & Real & $y=0$ & $\mathcal{O}(1/R)$\\
$\{A^{a}_{\mu}, \lambda^{p\,i}\}$ & $R^{3}_{V}[K_{\alpha}]$ & Real & $y=\pi R$ & $\mathcal{O}(1/R)$\\
$\{\bar{\lambda}^{\rho'\,i}, z^{\alpha'}\}$ & $R^{2}_{V}[K_{\alpha}]$ & Real & $y=\pi R$ & $\mathcal{O}(1/R)$\\\hline
\end{tabular} \end{center}
\textbf{Example}

Let's revisit the $SU(5,1)$ example based on the Lorentzian Jordan algebra $J^{\mathbb{C}}_{(1,5)}$.  We can obtain chiral multiplets (with a scalar potential) in the $\mathbf{(1,2)}\oplus \mathbf{(1,\bar{2})}$ of $SU(3)\times SU(2)\times U(1)$ (along with a spin-1/2 gauge singlet multiplet).  Let the indices correspond to:\\
\begin{center}\begin{tabular}{|c|c|c|c|c|}\hline
$I$ & $\alpha$ & $a'$ & $\alpha'$ & $0$\\ \hline
$SU(5,1)$ & $SM$ & $(1,2)\oplus (1,\bar{2})$ & $(3,1)\oplus(\bar{3},1)$ & (1,1)\\
 & \mbox{Gauge} & & $\oplus(3,2)\oplus (\bar{3},2)$ &  \\\hline
\end{tabular}\end{center}
 The $A^{\alpha}_{\mu}$ correspond to Standard Model gauge fields propagating on both fixed planes; the remaining vector fields either sit in massive multiplets, or are simply projected out.  In particular, we take the $A^{\alpha'}_{\mu}$ to be the $(3,2)\oplus (\bar{3},2)$ vectors (X, Y bosons) and color triplet vectors $(3,1)\oplus(\bar{3},1)$, which will propagate in massive supermultiplets in the effective theory of the $y=0$ plane.  This implies that massive spin-1/2 multiplets in the $[(3,2)\oplus (3,1)]\oplus [c.c.]$ will propagate in the effective theory of the $y=\pi R$ plane.  Next, let the $A^{a'}_{\mu}$ denote the vectors in the $(1,2)\oplus(1,\bar{2})$, which means there will be chiral multiplets in this representation at both fixed planes (with scalar potential terms).  Finally, we get conjugate pairs of massless chiral gauge singlet multiplets from the 5D supergravity multiplet.  There are no fields with index $a$ in this example.  

\subsection{Hypermultiplet sector}\label{sec:hypermultiplet sector 2}

So far, we have not been able to obtain massless chiral multiplets in \textit{complex} representations of the boundary gauge group.  Once again, the only way to do this (starting only from a 5D bulk theory) is to couple 5D hypermultiplets.  We can make an index split as in the previous cases:
\[
\tilde{X}=(X,X',\Omega,\Omega')\;\;\;\; A=(n,n',\tilde{n},\tilde{n}'),
\]    
where the fields have the following parity assignments under $\mathbb{Z}_{2}\times\mathbb{Z}_{2}$:\\
\begin{center}\begin{tabular}{|cccc|}
\hline
$++$ & $+-$ & $-+$ & $--$ \\ \hline
$q^{X}$ & $q^{X'}$ & $q^{\Omega}$ & $q^{\Omega'}$\\
$\xi^{n}_{1}$ & $\xi^{n'}_{1}$ & $\xi^{\tilde{n}}_{1}$ & $\xi^{\tilde{n}'}_{1}$\\
$\xi^{\tilde{n}}_{2}$ & $\xi^{\tilde{n}'}_{2}$ & $\xi^{n}_{2}$ & $\xi^{n'}_{2}$\\
\hline
\end{tabular}  \end{center}
The fields with $(+-)$ and $(-+)$ eigenvalues have massive $n=0$ modes, and so should fall into massive spin-1/2 multiplets.  Therefore, \textit{the indices $n$ and $\tilde{n}$ are required to be in ~1-1 correspondence as are the indices $n'$ and $\tilde{n}'$}.  However, there is no constraint between unprimed and primed indices, and each pair has an arbitrary range.  If $K$ is the 5D gauge group, and $K_{\alpha}$ is the boundary gauge group, the $K_{\alpha}$-representations of the massless chiral multiplets at the boundaries \textit{no longer need to form complete $K$-representations}.     

The multiplets with boundary propagating modes are listed below, along with some of their properties.  Start with $n_{H}$ 5D hypermultiplets in the $R_{H}[K]$ of the 5D gauge group $K$.  Let the gauge group at the orbifold fixed points be $K_{\alpha}$ so that under this group, the $R_{H}[K]$ decomposes into the representation 
\[ R_{H}[K_{\alpha}]=R^{1}_{H}[K_{\alpha}]\oplus R^{2}_{H}[K_{\alpha}], \]
where the indices $1$ and $2$ denote the splitting of $\tilde{X}$ into $(X,X')$.  At the fixed points, the hypermultiplets split into chiral multiplets with indices split into $(X,\Omega';X',\Omega)$, and are in the representations $R^{i}_{H}[K_{\alpha}]$ or $R^{i}_{H}[K_{\alpha}]\oplus \overline{R}^{i}_{H}[K_{\alpha}]$:
\begin{center}\begin{tabular}{|ccccc|}\hline
Multiplet & Representation & Type & Boundary & Mass\\\hline
$\{\xi^{n},q^{X_{1}}\}$ & $R^{1}_{H}[K_{\alpha}]$ & $\mathbb{R}$ or $\mathbb{C}$ & Both & Potential\\
$\{\xi^{\tilde{n}},q^{X_{2}}\}$ & $\overline{R}^{1}_{H}[K_{\alpha}]$ & $\mathbb{R}$ or $\mathbb{C}$ & Both & Potential\\
$\{\xi^{A'},q^{X'}\}$ & $R^{2}_{H}[K_{\alpha}]\oplus c.c.$ & $\mathbb{R}$ & $y=0$ & $\mathcal{O}(1/R)$\\
$\{\xi^{A'},q^{\Omega}\}$ & $R^{2}_{H}[K_{\alpha}]\oplus c.c.$ & $\mathbb{R}$ & $y=\pi R$ & $\mathcal{O}(1/R)$\\\hline
\end{tabular}\end{center}
We have split $X=(X_{1},X_{2})$ such that $X_{1}=1,\ldots, n_{H}$ and $X_{2}=n_{H}+1,\ldots, 2n_{H}$.  Also, $A'=(n',\tilde{n}')$ is a $USp(2m)$ index.\vspace{1mm}\\
\textbf{Example}

Consider the $SU(5)$ YMESGT with $C_{IJK}$ as in (\ref{canonical}) (where $C_{ijk}$ are the $d$-symbols of $SU(5)$), coupled to the minimal amount of Higgs and matter content in the bulk.  This can be realized by coupling the YMESGT to hypermultiplets whose scalars parametrize the quaternionic manifold~\footnote{By allowing an additional singlet hypermultiplet, we can instead couple the exceptional scalar manifold $\frac{E_{8}}{SU(6)\times SU(2)}$.}
\[ 	
\mathcal{M}_{Q}=\frac{SU(27n,2)}{SU(27n)\times SU(2)\times U(1)},
\] 
resulting in a coupling of $n$ sets of hypermultiplets in the $\mathbf{1}\oplus 3(\mathbf{5})\oplus \mathbf{10}$ of $SU(5)$~\cite{M05a}.  

Suppose we are going to break $SU(5)\rightarrow SU(3)\times SU(2)\times U(1)$; focussing on the hypermultiplet sector, we can make the following assignments 
\begin{center}\begin{tabular}{|c|c|c|c|}\hline
$n$ & $\tilde{n}$ & $n'$ & $\tilde{n}'$\\\hline
$\,Matter \oplus (1,2)\oplus (1,\bar{2})\,$ & $\,c.c.\,$ & $\,(3,1)\oplus (\bar{3},1)\,$ & $\,c.c.\,$\\\hline  
\end{tabular}\end{center}   
This will result in a low energy theory at both boundaries with Standard Model chiral matter multiplets; a pair of left-chiral Higgs doublets and their CPT conjugates; and a pair of massive spin-1/2 color triplet multiplets, all at both boundaries.  

\section{Conclusion}\label{sec:conclusion}

We have found parity assignments for fields and other objects in five-dimensional Yang-Mills-Einstein supergravity coupled to tensor and hypermultiplets on $M_{4}\times S^{1}/\Gamma$, allowing for general gauge symmetry breaking.  We have used the dimensionally reduced Lagrangians, truncating our attention to the zero mode sector of the theory, though the parity assignments are true for the full supergravity theory.  

For $\Gamma=\mathbb{Z}_{2}$, the generic result is that bulk gauge multiplets for a group $K$ in the bulk yield boundary gauge multiplets for the remaining group $K_{\alpha}$, and chiral multiplets in $K/K_{\alpha}$ forming a real representation of $K_{\alpha}$; bulk hypermultiplets yield chiral multiplets on the boundary forming either real or complex $K_{\alpha}$-reps; and bulk tensor multiplets may yield either chiral multiplets in real $K_{\alpha}$-reps or massive charged vector multiplets, depending on the parity for the components of the associated symplectic metric.  The boundary theories involving massive vector multiplets charged under $K_{\alpha}$ are the analogues of the $\mathcal{N}=2$ dimensionally reduced theories in~\cite{GMZ05b}.

The extension to the case $\Gamma=\mathbb{Z}_{2}\times\mathbb{Z}_{2}$ allows for light boundary chiral multiplets in incomplete $K$-representations \textit{except} those coming from 5D tensor multiplets.  This is the usual mechanism for `doublet-triplet' splitting in orbifold-GUTs.    

A novel feature of supergravity theories is that one can gauge a non-compact group.  The vectors representing the non-compact generators must be given odd parity, yielding a 4D compact gauge group and chiral multiplets in real representations.     

While not examined here, the classical scalar potential on the boundaries can be determined from the form of the parity assignments and the dimensionally reduced Lagrangians, which are available in the literature quoted.  There is a positive-definite contribution from the 5D pure YMESGT and tensor sector, while hypermultiplet couplings can contribute negative scalar potential terms.  The potentials from compact and non-compact gauge multiplets are distinct.  The overall potential can then be analyzed for groundstates spontaneously breaking supersymmetry or electroweak symmetry. 

Although boundary-localized fields are not presumed, assigning even parity to 4D $\mathcal{N}=1$ vector multiplets requires the moduli space of the boundary theory to lie on the boundary of the clasical K\"{a}hler cone.  Therefore, a consistent supergravity description requires boundary-localized fields to be added, which in a compactified 11D supergravity description correspond to membrane states that become massless when wrapping collapsed Calabi-Yau 2-cycles (the collapse occuring over the fixed points of the effective 5D theory).  In particular, gauge symmetry enhancement may occur at these points. 

\section{Future Directions}\label{sec:future directions}
We will examine some issues regarding symmetries, anomalies, and the presence of the singlet scalars appearing in the orbifold theory.  There is always a QCD-type axion present in 5D supergravity on $S^{1}/\Gamma$, and we will describe the form of the couplings.

Now that parities of various objects appearing in supergravity have been listed, one can examine the terms in the Lagrangian, including $\partial_{5}\Phi$ terms and some of the fermion interactions that were mostly neglected here.  In the upstairs picture, as in the case of simple supergravity~\cite{JB}, this will result in a bulk Lagrangian along with terms supported only at the orbifold fixed points.  In the downstairs picture, this simply corresponds to a 5D Lagrangian with boundary conditions on the fields.  This will allow us to examine the supersymmetry transformation laws for the fields in the presence of orbifold fixed points, which will lead to a study of global solutions to the Killing spinor equations.  Much work has been done using such an analysis in the case of gauged simple supergravity (see e.g.~\cite{JB} and references therein), mostly in the context of braneworld scenarios.  \vspace{4mm}\\  
\textit{\textbf{Acknowledgements}}\\
The author would like to thank Jonathan Bagger for enlightening conversation, and Murat G\"{u}naydin for helpful comments regarding the writing of the manuscript.\vspace{8mm}\\  
\textbf{\Large{Appendices}}
\appendix
\section{Notation and Conventions}\label{sec:notation and conventions}

We use the following notation for representations of Lie groups.  Consider a set of $m$ 5D $\mathcal{N}=2$ hypermultiplets, which contain $4m$ real scalars that form the $(\mathbf{m}\oplus\mathbf{\bar{m}})\oplus(\mathbf{m}\oplus\mathbf{\bar{m}})$ of a group $G$ (assuming that we are not dealing with pseudoreal representations), where the dimension of the representation is real.  Since the scalars form $m$ quaternions, the representation can be written as $\mathbf{m_\mathbb{H}}$.  We will simply say that \textit{the $m$ hypermultiplets are in the $\mathbf{m}$ of $G$}, dropping the subscript.  On the other hand, consider $n$ 4D $\mathcal{N}=1$ left-chiral multiplets in the $\mathbf{n}$ of $G$ and their $n$ right-chiral conjugate partners in the complex conjugate $\mathbf{\bar{n}}$ of $G$, where again we use the real dimension of the representation.  If the chiral multiplets all combine with their conjugates to form massive spin-1/2 multiplets, we will use the complex dimension $n_{\mathbb{C}}$ and simply say \textit{the spin-1/2 multiplets are in the $\mathbf{n}$ of $G$}, dropping the subscript.        

We use the signature $\eta_{mn}=diag(-,+,+,+,+)$.  
For gamma matrices, we take
\[ \Gamma^{\mu} = \left( \begin{array}{ccc}
0 & & \sigma^{m} \\ -\sigma^{m} & & 0
\end{array} \right) \;\;\;\;
\Gamma^{5}= \left( \begin{array}{ccc}
i & & 0 \\
0 & & -i
\end{array} \right) \]   
where $\sigma^{m}$ are the spacetime Pauli matrices, and $m$ is a flat spacetime index.  The charge conjugation matrix is taken to be 
\[C=\left(
\begin{array}{ccc}
e & & 0\\
0 & & -e
\end{array}
\right)\;\;\;\;\;\mbox{where}\;\;\;\;\;
e=\left(
\begin{array}{ccc}
0 & & -1\\
1 & & 0
\end{array}
\right).\] 
The charge conjugation matrix therefore satisfies 
\[C^{T}=-C=C^{-1}\;\;\;\;\;\mbox{and}\;\;\;\;\; C\Gamma^{m}C^{-1}=(\Gamma^{m})^{T}.\]
In five spacetime dimensions, there are a minimum of eight supercharges; since there is a global $SU(2)_{R}$ symmetry, it is convenient to use symplectic-Majorana spinors, which form an explicit $SU(2)_{R}$ doublet.  Given a 4-component spinor $\lambda$, the Dirac conjugate is defined by  
\[\bar{\lambda}^{i}=(\lambda_{i})^{\dagger}\Gamma^{0},\]
where $i$ is an $SU(2)_{R}$ index, which is raised and lowered according to
\[\lambda^{i}=\epsilon^{ij}\lambda_{j},\;\;\;\;\lambda_{j}=\lambda^{i}\epsilon_{ij},\]
with $\epsilon_{12}=\epsilon^{12}=1$.  Then a symplectic-Majorana spinor is one that satisfies 
\[\bar{\lambda}^{i}=\lambda^{i\,T}C.\]
We will take the following form for our Majorana spinors showing the 2-component spinor content:
\[
\lambda^{1}=\left(
\begin{array}{c}
\xi\\
e\zeta^{*}
\end{array}
\right)\;\;\;\;\lambda^{2}=\left(
\begin{array}{c}
\zeta\\
-e\xi^{*}
\end{array}
\right).
\]

\section{Parity assignments for fermionic fields}\label{sec:parity assignments for fermionic fields}

The parity assignments for the components of the symplectic Majorana spinors in eqn~(\ref{5dfermions}) and~(\ref{5dfermions2}) are listed here.  Some symplectic pairs $\lambda^{1},\lambda^{2}$ become left/right chiral spinors on the spacetime boundaries, while others denoted in the paper as $\bar{\lambda}^{1},\bar{\lambda}^{2}$ become right/left chiral spinors.
\begin{center}
\begin{tabular}{|c|c|}  \hline
Even & Odd \\ \hline 
$\alpha_{\mu}\;\;\;\;\alpha^{*}_{\mu}$ & $\beta_{\mu}\;\;\;\;\beta^{*}_{\mu}$\\ 
$\beta_{\dot{5}}\;\;\;\;\beta^{*}_{\dot{5}}$ & $\alpha_{\dot{5}}\;\;\;\;\alpha^{*}_{\dot{5}}$\\
$\gamma^{p}\;\;\;\;\gamma^{p\,*}$ & $\delta^{p}\;\;\;\;\delta^{p\,*}$\\
$\delta^{\rho}\;\;\;\;\delta^{\rho\,*}$ & $\gamma^{\rho}\;\;\;\;\gamma^{\rho\,*}$\\
$\eta\;\;\;\;\eta^{*}$ & $\zeta\;\;\;\;\zeta^{*}$ \\ 
$\zeta^{n}_{1}\;\;\;\;\zeta^{\tilde{n}}_{2}$ & $\zeta^{n}_{2}\;\;\;\;\zeta^{\tilde{n}}_{1}$\\
\hline \end{tabular}
\end{center}

\end{document}